\documentclass[aps,superscriptaddress,twocolumn]{revtex4}

\usepackage{mathptmx}
\usepackage{sublabel}
\usepackage{dcolumn}
\usepackage{subfigure}
\usepackage{amsmath}
\usepackage{amssymb}
\usepackage{amsfonts}
\usepackage{bm}
\usepackage{bbm}
\usepackage{color}
\usepackage{overpic}
\usepackage{latexsym}
\usepackage{epstopdf}
\usepackage{color}
\usepackage[english]{babel}
\usepackage{latexsym}
\usepackage{stmaryrd}

\usepackage{psfrag,graphicx}
\usepackage{epsf}
\usepackage{natbib}
\usepackage{epstopdf}
\usepackage{appendix}
\usepackage{dsfont}
\usepackage{mathtools}

\definecolor{mygrey}{gray}{0.35}
\definecolor{myblue}{rgb}{0.2,0.2,0.8}
\definecolor{myzard}{cmyk}{0,0,0.05,0}
\definecolor{mywhite}{rgb}{1,1,1}
\definecolor{myred}{rgb}{1,0.,0.3}

\usepackage[colorlinks=true,citecolor=myblue,linkcolor=myred]{hyperref}
\def\beq{\begin{equation}}
\def\eeq{\end{equation}}
\def\ba{\begin{align}}
\def\enda{\end{align}}
\def\bi{\begin{itemize}}
\def\ei{\end{itemize}}

\newcommand{\eq}[1]{Eq.\;(\ref{#1})}

 \def\ee{\mathord{\rm e}}
 
 \def\ii{\mathord{\rm i}}

 \def\ee{\mathord{\rm e}}
 
 \def\ii{\mathord{\rm i}}

\renewcommand{\ii}{{\rm i}}
\renewcommand{\ee}{{\rm e}}

 \newcommand{\ket}[1]{|#1\rangle}
 \newcommand{\bra}[1]{\langle #1|}

 \newcommand{\ketbradif}[2]{\ket{#1}\bra{#2}}
 \newcommand{\ketbra}[1]{\ketbradif {#1}{#1}}

\graphicspath{{Plots/}}

\begin{document}

\title[Short Title]{Simulating spin-boson models with trapped ions}

\author{A. Lemmer}
\email{andreas.lemmer@uni-ulm.de}
\affiliation{Institut f\"ur Theoretische Physik and $IQ^{ST}$, Universit\"at Ulm, Albert-Einstein Alle 11,
89069 Ulm, Germany}

\author{C. Cormick}
\affiliation{IFEG, CONICET and Universidad Nacional de C\'ordoba, X5000HUA, C\'ordoba, Argentina}

\author{D. Tamascelli}
\affiliation{Institut f\"ur Theoretische Physik and $IQ^{ST}$, Universit\"at Ulm, Albert-Einstein Alle 11,
89069 Ulm, Germany}
\affiliation{Dipartimento di Fisica, Universit\`a  degli Studi di Milano, Via Celoria 16, 20133 Milano, Italy}

\author{T. Schaetz}
\affiliation{Physikalisches Institut, Albert-Ludwigs-Universit\"at Freiburg,
Hermann-Herder-Str.3, 79104 Freiburg, Germany}

\author{S. F. Huelga}
\affiliation{Institut f\"ur Theoretische Physik and $IQ^{ST}$, Universit\"at Ulm, Albert-Einstein Alle 11,
89069 Ulm, Germany}

\author{M. B. Plenio}
\email{martin.plenio@uni-ulm.de}
\affiliation{Institut f\"ur Theoretische Physik and $IQ^{ST}$, Universit\"at Ulm, Albert-Einstein Alle 11,
89069 Ulm, Germany}

\begin{abstract}
We propose a method to simulate the dynamics of spin-boson models with small crystals of trapped ions where the electronic degree of freedom of one ion is used to encode the
spin while the collective vibrational degrees of freedom are employed to form an effective harmonic environment. The key idea of our approach is that a single damped mode can be used to
provide a harmonic environment with Lorentzian spectral density. More complex spectral functions can be tailored by combining several individually damped modes. We propose to work with mixed-species crystals such that one species serves to encode the spin while the
other species is used to cool the vibrational degrees of freedom to engineer the environment. The strength of the dissipation on the spin can be controlled by tuning the coupling between spin and vibrational degrees of freedom. In this way the dynamics
of spin-boson models with macroscopic and non-Markovian environments can be simulated using only a few ions. We illustrate the approach by simulating an experiment with
realistic parameters and show by computing quantitative measures that the dynamics is genuinely non-Markovian.
\end{abstract}

\date{\today}
\maketitle

The spin-boson model is an archetypical model of an open quantum system with applications ranging from chemical reactions~\cite{garg_lorentzian_sd} over biological aggregates \cite{dong_schulten} to solid state physics~\cite{weiss_book,leggett_review, bulla_review}.
The model describes a single spin coupled to a dissipative environment comprised by an infinite set of harmonic oscillators.
It is well known that the effect of thermal oscillator environments on a quantum system is fully described by a single scalar function, the spectral density (or spectral
function) of the environment~\cite{leggett_review}. Although approximate analytic solutions have been found for some spectral densities~\cite{weiss_book, leggett_review} no closed analytic solution of the spin-boson model is known.
Meanwhile, dynamics and thermodynamical properties of spin-boson models have been investigated by a number of numerical approaches including techniques based on the numerical renormalization 
group~\cite{bulla_review}, time-dependent density matrix renormalization group~\cite{chain_mapping, prior}, path integral Monte Carlo~\cite{pimc},
or the quasi-adiabatic propagator path integral approach~\cite{quapi}. Numerical simulations are especially needed for environments with spectral densities 
where the reorganization energy is of the order of the spectral width or highly structured environments with long-lived vibrational modes that lead to highly non-trivial dynamics.
These types of spectral densities are of particular relevance for the excitonic and electronic dynamics in biomolecular systems~\cite{huelga_plenio_contemp_physics}
and pose considerable challenges for numerical methods especially when the results of non-linear spectroscopy need to be predicted~\cite{almeida_j_chem_phys}. Therefore, an experimental simulator with a high degree of control is desirable.

Trapped atomic ions provide a clean and highly controllable system where many dynamical quantities are directly accessible. They have proven 
to be a versatile platform for the simulation of a wide range of physical models, such as defect formation in classical phase transitions~\cite{kibble_zurek_0, kibble_zurek_1, kibble_zurek_2} 
as well as open and closed quantum systems~\cite{qsim_review, manybody_qsim, thermalization_schaetz, blatt_oqs, haeffner_correlations, monroe_mbl}. 
The simulation of spin-boson models using trapped atomic ions has been proposed
previously~\cite{porras_sbm} requiring rather large crystals comprising 50-100 ions. Such crystals feature a large number of vibrational modes which can be used to act as a mesoscopic
environment for the spin. However, for these large crystals the level of control needed to simulate spin-boson models is not available in the foreseeable future. 
In this work, we develop a proposal to simulate the dynamics of spin-boson models using small crystals of trapped ions.
Our procedure also relies on the vibrational degrees of freedom to model the environment, but it makes use of the fact that a damped mode produces an effective Lorentzian 
spectral density~\cite{garg_lorentzian_sd}. While in~\cite{garg_lorentzian_sd} the damping is provided by an oscillator reservoir with Ohmic spectral density we show that the same spectral density
can be obtained in certain regions of parameter space if the damping is modeled by a Lindblad equation extending the results of~\cite{imamoglu1, imamoglu2,garraway1, garraway2}.
The resulting spectral densities are continuous functions of frequency and can thus be identified with an environment made up of a macroscopic number of modes as it occurs in the condensed phase.
Controlling the couplings of the spin to the modes, the mode frequencies and the damping rates, the shape of the spectral density can be tailored, allowing one to mimic environments with continuous
and highly-structured spectral densities using only a small number of oscillators to form the environment. This reduced overhead brings the simulation of spin-boson models to the realm of state-of-the-art trapped-ion setups. 

{\it Spin-boson model.--} The spin-boson model describes a two-level system (spin 1/2) in a dissipative environment which is modeled by an infinite set of non-interacting harmonic oscillators.
Denoting by $\epsilon$ the energy splitting between the spin states $\ket{\uparrow}$ and $\ket{\downarrow}$ and by $\hbar \Delta$ the coupling between them, the Hamiltonian of the global system reads~\cite{weiss_book}
\beq
H_{\rm sb} =\frac{\epsilon}{2}\sigma^z - \frac{\hbar \Delta}{2}\sigma^x - \frac{1}{2}\sigma^z \sum_{n} \hbar \lambda_{n} (a_{n} + a^{\dagger}_{n}) + \sum_{n} \hbar \omega_{n} a^{\dagger}_{n} a_{n}
\label{sb_ham_main}
\eeq
where $\sigma^{z}= \ketbra{\uparrow}-\ketbra{\downarrow}$ and $\sigma^{x}= \ketbradif{\uparrow}{\downarrow} + \ketbradif{\downarrow}{\uparrow}$. $a_{n}^{\dagger} (a_{n})$ denotes the raising (lowering) operator of
environmental mode $n$ and $\omega_{n}$ the corresponding frequency while the real $\lambda_{n}$ describe the couplings of the spin to the environmental oscillators. The spectral density which determines the influence 
of the oscillator environment on the spin~\cite{leggett_review} reads $ J(\omega) = \pi \sum_n \lambda_n^2 \delta (\omega -\omega_n)$ with $\delta$ the Dirac $\delta$-function. For a macroscopic environment one assumes 
that the frequencies are so closely spaced that $J(\omega)$ becomes a continuous function of $\omega$.

One is generally interested in finding the reduced dynamics of the spin for an environment with a certain spectral density.
The path-integral formalism~\cite{feynman_hibbs} provides us with an exact expression for the propagator of the spin state where the effects of the environment are already included. For factorizing initial conditions 
$\rho_0 = \rho_s \otimes \rho_{\beta}$ with some spin state $\rho_s$ and the environmental modes in a thermal state $\rho_\beta$ at inverse temperature
$\beta = (k_B T)^{-1}$ the propagator for the spin reads~\cite{feynman_vernon}
\beq
G (t,0) = \int_{q_0} ^{q_f} D q \int_{q'_0} ^{q'_f} D q' \ee^{\frac{\ii}{\hbar}(S_0[q]-S_0[q'])} F[q,q'].
\label{pi_propagator}
\eeq
Here the path integral $\int_{q_0} ^{q_f} D q$ runs over all spin state trajectories connecting $q(0)=q_0$ and $q(t) =q_f$, $S_0[q]$ is the action of the free spin evolution
and $F [q,q']$ is the Feynman-Vernon influence functional~\cite{feynman_vernon}. The influence functional contains the effect of the environment on the spin dynamics.
For an oscillator environment and the considered coupling it can be written as~\cite{weiss_book} 
\beq
\begin{split}
F [q,q'] = & \exp\left\{ -\int_0^{t} {\rm d} t' \int_{0}^{t'} {\rm d} s [q(t') - q'(t')] \right. \\
&\left. [L(t'-s) q(s) - L^* (t'-s) q'(s)] \right\}
\end{split}
\label{fv_influence_phase}
\eeq
where $L(t) = \frac{1}{\hbar^2} \langle X(t) X(0) \rangle_{\beta}$ is the reservoir correlation function with $X= \sum_n \hbar \lambda_n (a_n + a_n^{\dagger})$.
Alternatively, $L(t)$ can be expressed in terms of the spectral density $J(\omega)$:
\beq
L(t) = \frac{1}{\pi} \int_0^{\infty} {\rm d}\omega\, J(\omega) \left[\coth\left(\frac{\beta \hbar \omega}{2}\right) \cos (\omega t) - \ii \sin(\omega t)\right].
\label{autocorr_sd}
\eeq

{\it Spectral density of damped harmonic oscillators.--} The key idea of our approach is the fact that a damped oscillator provides a continuous effective spectral density,
and the observation that different environments that produce the same influence functional have the same effect on the spin dynamics~\cite{feynman_vernon}. 

Let us first consider an environment consisting of a single harmonic oscillator which is damped by an oscillator reservoir with Ohmic spectral function.
If we denote the free oscillation frequency of the damped oscillator by $\Omega$ and the bath causes damping at rate $\kappa$ on the oscillator, the effective spectral density
generated by the damped oscillator on the spin is Lorentzian~\cite{garg_lorentzian_sd, supp_mat}
\beq
J_{\rm eff} (\omega) = \lambda^2 \left[ \frac{\kappa}{\kappa^2 +(\omega-\omega_{\rm m})^2} - \frac{\kappa}{\kappa^2 +(\omega+\omega_{\rm m})^2}\right].
\label{jeff_main}
\eeq
Here $\omega_{\rm m} = \sqrt{\Omega^2 - \kappa^2}$ is the reduced frequency of the damped oscillator and $\hbar \lambda$ the spin-oscillator coupling as in~\eq{sb_ham_main}. Note that we 
restrict our considerations to the underdamped regime $\kappa < \Omega$. 

The combined influence functional of several independent damped harmonic oscillators is given by the product of the individual influence functionals~\cite{feynman_vernon}.
Therefore, if the reservoirs have the same temperature, according to Eqs.~\eqref{fv_influence_phase} and~\eqref{autocorr_sd} their spectral densities add up  and one can construct effective
spectral densities $J (\omega) = \sum_n J_{{\rm eff},n} (\omega).$
Here $J_{{\rm eff},n} (\omega)$ is the spectral density due to oscillator $n$ given by~\eq{jeff_main} with the corresponding $\lambda_n, \kappa_n, \omega_n$. If one wants to approximate
a certain target spectral density $J_{\rm t}(\omega)$ the values for $\lambda_n, \kappa_n, \omega_n$ are found by minimizing the functional $
E[ \{ \lambda_n, \kappa_n, \omega_n \} ] = \int_0^{\infty} {\rm d}\omega |J_{\rm t}(\omega) - J(\omega)|^2$ as has been shown in~\cite{tannor_parametrization}.

In trapped-ion experiments, the motion of the ions is usually expressed in terms of a set of normal modes, each of which is a harmonic oscillator. Cooling of the modes
is commonly described by a Lindblad equation~\cite{cirac_laser_cooling, morigi_eit_cooling}. Therefore, it is not immediately clear if we can obtain an effective spectral density as
for the oscillator damped by an Ohmic bath,~\eq{jeff_main}. We will now show that this is possible and we obtain the same spectral function for appropriate parameters.

Let us start by considering the reservoir correlation function $L(t)$ in~\eq{autocorr_sd}. We note that $L(t) =  L'(t) + \ii L'' (t) $ is a complex-valued function with real
and imaginary parts $L'(t)$ and $L''(t)$. For the oscillator damped by an Ohmic bath the coordinate correlation function and thus $L(t)$ can be calculated
analytically~\cite{grabert_damped_qho, weiss_book, supp_mat} and we obtain $L' (t) = L_1(t) + L_2(t)$
\beq
\begin{split}
L_1(t) &= \lambda^2 \left[\frac{\sinh(\beta \hbar \omega_{\rm m})}{\cosh(\beta \hbar \omega_{\rm m}) -\cos(\hbar \beta \kappa)} \cos(\omega_{\rm m} t) \right.  \\
& \left. +\frac{\sin(\hbar \beta \kappa)}{\cosh(\beta \hbar \omega_{\rm m}) -\cos(\hbar \beta \kappa)} \sin(\omega_{\rm m} |t|) \right]\ee^{-\kappa |t|},
 \\ 
L_2(t) &= -\lambda^2 \frac{8 \kappa \omega_{\rm m}}{\hbar \beta} \sum_{n=1}^{\infty} \frac{\nu_n \ee^{-\nu_n |t|}}{(\Omega^2 + \nu_n^2)^2 - 4\kappa^2 \nu_n^2} \\
\end{split}
\label{l_real}
\eeq
with the Matsubara frequencies $\nu_n = 2 \pi n/(\hbar \beta)$  and
\beq
L'' (t) = -\lambda^2 \sin(\omega_{\rm m} t) \ee^{-\kappa |t|}.
\label{l_imag}
\eeq

In Lindblad description, a damped harmonic oscillator coupled to a thermal reservoir at inverse temperature $\beta$ evolves according to
\beq
\dot{\rho} = -\frac{\ii}{\hbar} [H,\rho] + \mathcal{D}_{\kappa,\bar{n}} \rho 
\label{lindblad_main}
\eeq
where here $H = \hbar \omega_{\rm m} a^{\dagger} a$ is the Hamiltonian of the oscillator and its frequency $\omega_{\rm m}$ already includes possible renormalizations due to the damping.
The dissipator reads~\cite{breuer_petruccione}
\beq
\mathcal{D}_{\kappa,\bar{n}} \rho = \kappa (\bar{n} +1) [a \rho a^{\dagger} - a^{\dagger} a \rho] + \kappa \bar{n} [a^{\dagger} \rho  a - a a^{\dagger} \rho  ] + {\rm H.c.}
\label{lindblad_dissipator_main}
\eeq
Using the quantum regression theorem we can obtain the reservoir correlation function $L_{\rm L} (t) = L_{\rm L}' (t) + \ii L_{\rm L}'' (t)$
for the damped harmonic oscillator in Lindblad description. We find that the real part
\beq
L'_{\rm L} (t) = \lambda^2 \coth\left(\frac{\beta \hbar \omega_{\rm m}}{2}\right) \,\cos(\omega_{\rm m} t) \ee^{-\kappa |t|}
\label{l1l_main}
\eeq
has a different functional form than $L' (t)$ in~\eq{l_real} while the imaginary part $L_{\rm L}'' (t)$ coincides with $L '' (t)$ in~\eq{l_imag} which is determined by $J_{\rm eff} (\omega)$ of~\eq{jeff_main}.
Writing $L_{\rm L}'(t)$ as in~\eq{autocorr_sd} we obtain $L_{\rm L}' (t) =  \frac{1}{\pi} \int_0^{\infty} {\rm d}\omega\, \tilde{J}_{\rm eff} (\omega) \coth (\beta \hbar \omega/2) \cos (\omega t) $ where 
\beq
\tilde{J}_{\rm eff} (\omega) = \lambda^2 \frac{\coth\left(\frac{\beta \hbar \omega_{\rm m}}{2}\right)}{\coth\left(\frac{\beta \hbar \omega}{2}\right) } \left[\frac{\kappa}{\kappa^2+(\omega-\omega_{\rm m})^2} + \frac{\kappa}{\kappa^2+(\omega+\omega_{\rm m})^2}\right].
\label{jeff_tilde_main}
\eeq

Despite the differences it is possible to obtain a very good agreement between the real parts $L' (t)$ and $L_{\rm L}' (t)$ and their frequency space representations Eqs.~\eqref{jeff_main} and~\eqref{jeff_tilde_main}. 
Ref.~\cite{talkner_fdt} estimates that the quantum regression theorem can only yield quantitatively correct predictions for the two-time correlation functions of the damped harmonic oscillator if $\kappa \ll \omega_{\rm m}$ and $\hbar \beta \kappa \ll 1$.
Indeed, under these assumptions we find very good agreement between $L' (t)$ and $L_{\rm L}' (t)$. If we have good agreement between $L' (t)$ and $L_{\rm L}' (t)$, we also find good agreement in frequency space.  Note that while $\kappa \ll \omega_{\rm m}$ is a necessary
condition to derive the Lindblad equation~\eqref{lindblad_main} with the dissipator in~\eq{lindblad_dissipator_main}, $\hbar \beta \kappa \ll 1$ puts a lower bound on the temperature where the identification of $L (t)$ and $L_{\rm L} (t)$  is possible. On the other hand, 
also too high temperatures lead to deviations such that there is an intermediate temperature range where the best agreement is achieved (see~\cite{supp_mat} for a more detailed discussion).

In order to confirm the above considerations we simulated the dynamics of $\langle \sigma^z (t) \rangle$ for the full spin-boson Hamiltonian in~\eq{sb_ham_main} with spectral density $J_{\rm eff}(\omega)$ from~\eq{jeff_main} using the numerically exact
TEDOPA algorithm~\cite{chain_mapping} and compared them with those given by~\eq{lindblad_main} with $H = H_{\rm sb}$ from~\eq{sb_ham_main} for a single mode. We considered an initial product state $\ketbra{\uparrow} \otimes \rho_{\beta}$
and $\epsilon = 0$, $\omega_{\rm m}/2\pi = 100\,$kHz, $\kappa/2\pi = 1.25\,$kHz as well as a spin-mode coupling $\lambda/2\pi=100\,$kHz. We chose $\hbar \beta = 5.91\cdot 10^{-6}\,{\rm s}$ which corresponds to $\bar{n}(\omega_{\rm m}) = 0.025$ for the Lindblad-damped oscillator and
computed the evolution for spin energies $\Delta/2\pi=50\,$kHz and $100\,$kHz. For both values of $\Delta$ we obtain very good agreement (see~\cite{supp_mat}) which shows that the analogy to the macroscopic environment also holds when we probe the spectral density
away from the resonance. Note that one simulation for $\Delta/2\pi=50\,$kHz takes 15 days using 16 cores on a computing cluster which once more indicates the value of a trapped-ion simulator especially for structured environments and complex observables.

\begin{figure*}[hbt]
\centering
\includegraphics[width=.35\textwidth]{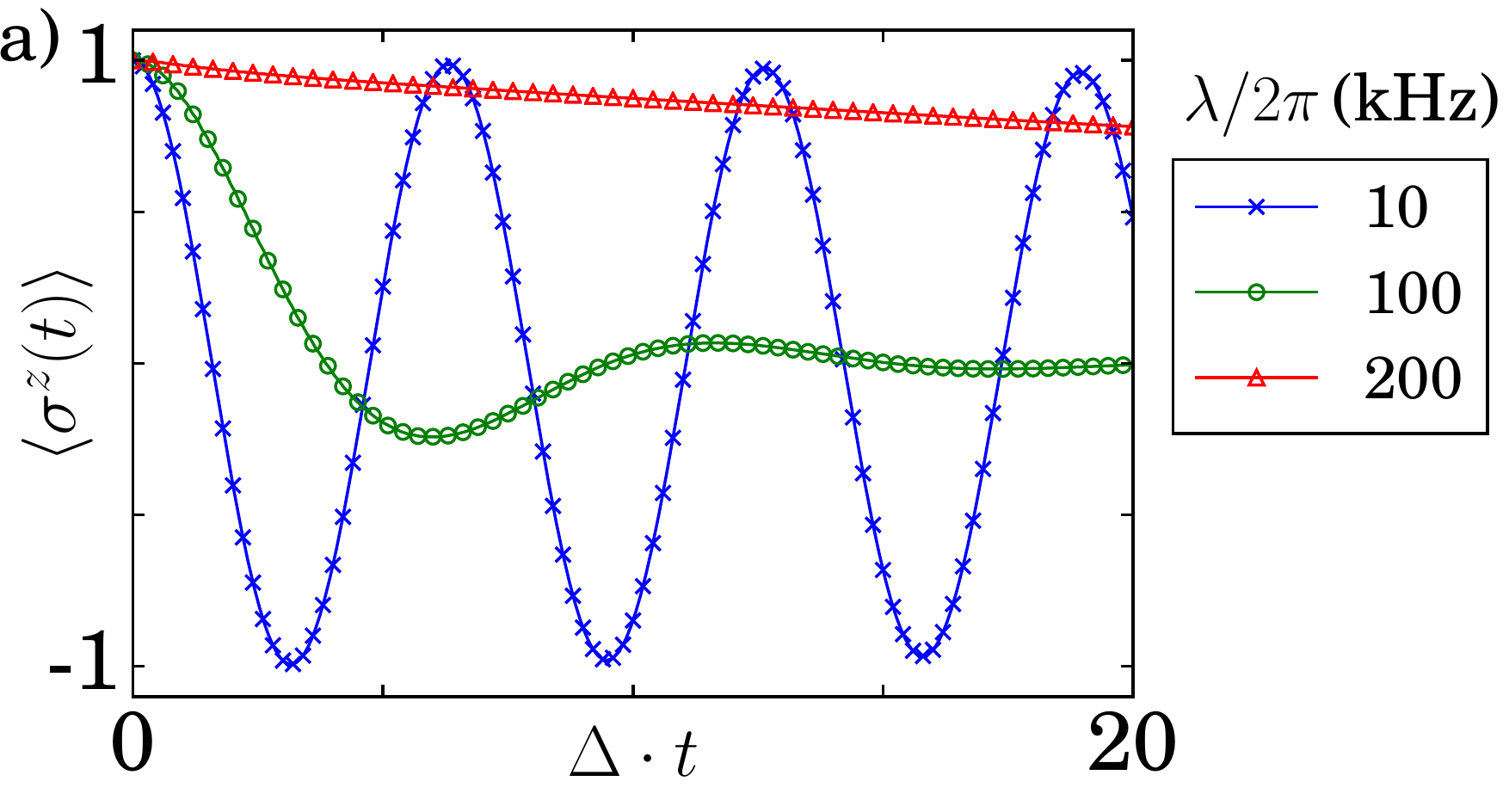}\hfill
\includegraphics[width=.31\textwidth]{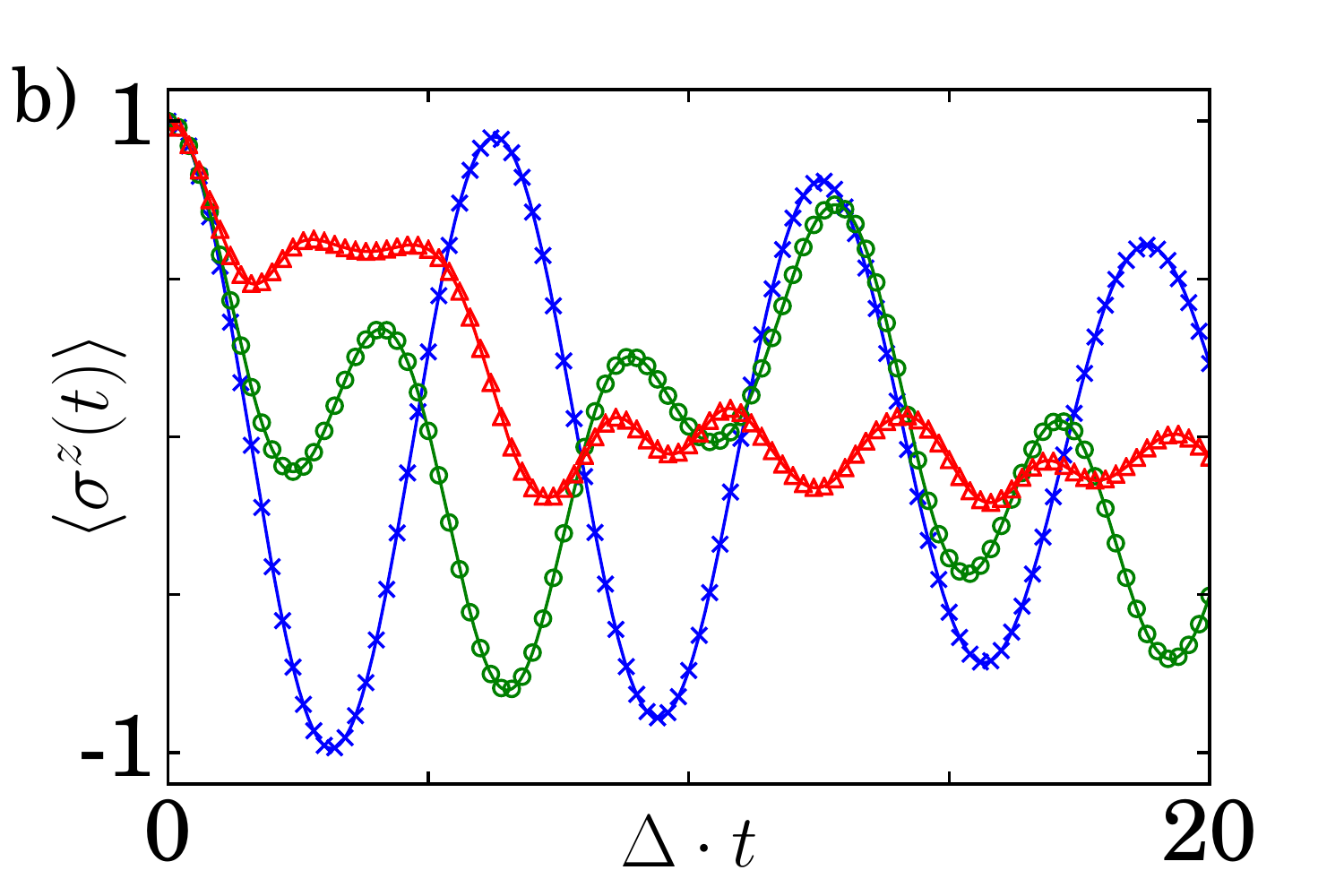}\hfill
\includegraphics[width=.32\textwidth]{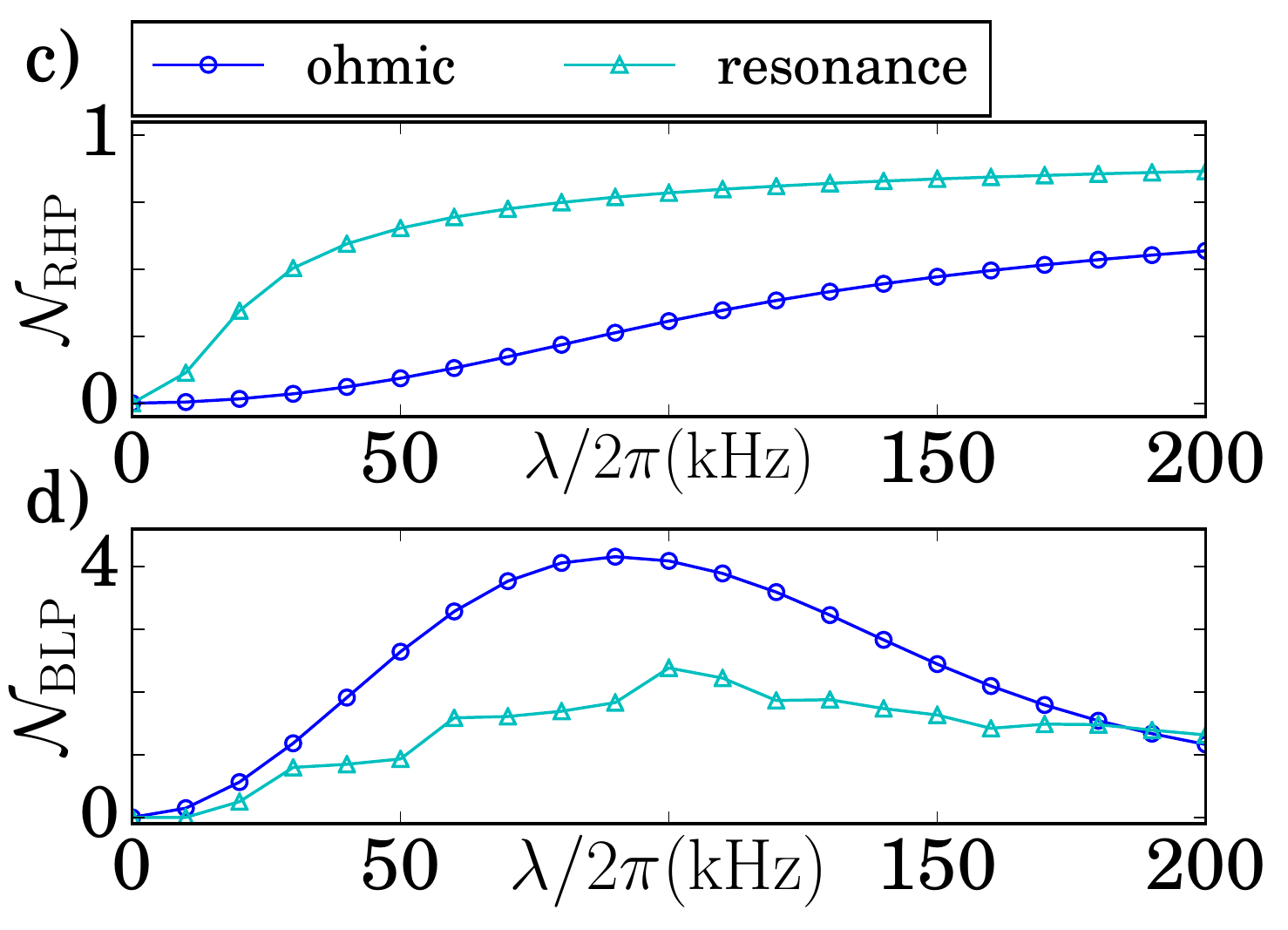}
\caption{Parts {\bf a)} and {\bf b)} of the figure show the dynamics of $\langle \sigma_z (t) \rangle$ in natural time units $\Delta \cdot t$ under~\eq{lindblad_main} with $H=H_{\rm sb1}$ from~\eq{sb_ham_ions_main}, which corresponds to a spin-boson model
with a Lorentzian spectral density as in~\eq{jeff_main}, for varying spin-motion coupling $\lambda$. In part{ \bf a)} the spin energy $\Delta/2\pi = 3\,$kHz is much smaller than the mode frequency $\omega_{\rm m}/2\pi = 100\,$kHz, so that the environment is approximately Ohmic. In part {\bf b)} the mode 
is resonant with the spin ($\Delta/2\pi = 100\,$kHz). The remaining parameters are given in the text. Part {\bf c)} shows the measure of non-Markovianity $\mathcal{N}_{\rm RHP}$ in the intervals $[0.01/\Delta]$ and $[0.1/\Delta]$ for the ohmic and resonant cases, respectively.
Part {\bf d)} depicts the measure of non-Markovianity $\mathcal{N}_{\rm BLP}$ over the whole interval $[0,20/\Delta]$ for both cases. }
\label{sb_dynamics}
\end{figure*}

{\it Trapped ion implementation. --} Let us now proceed to illustrate how the ideas discussed above can be implemented in an ion-trap experiment. We consider $N$ singly charged atomic ions with masses $m_j$ confined
in a linear Paul trap with effective harmonic trapping potential. We assume trapping conditions such that laser cooled ions form a linear Coulomb crystal along $z$ with equilibrium positions ${\bf r}_j^0=(0,0,z_j^0)^T$.
The motional degrees of freedom can then be described in terms of $N$ uncoupled normal modes in each spatial direction~\cite{james_hom_crystals, morigi_inhom_crystals}
and the motional Hamiltonian reads $H_m = \sum_{n,\alpha} \hbar \omega_{n,\alpha} a_{n,\alpha}^{\dagger} a_{n,\alpha}$ where $\omega_{n,\alpha}$ is the frequency of mode $n$ in spatial direction $\alpha\in\{x,y,z\}$ with
ladder operators $a_{n,\alpha}^{\dagger},\: a_{n,\alpha}$.

For simplicity, we will focus on the case of a spin coupled to a single damped mode which corresponds to a spin-boson model with Lorentzian spectral density as in~\eq{jeff_main}. This system already exhibits an interesting
phenomenology and has been studied with a variety of numerical and analytical approaches, see e.g.~\cite{mth_lorentzian_sd, fkw_lorentzian_sd, ankerhold_lorentzian_sd, caldeira_lorentzian_sd}. For this purpose we only need $N=2$ ions: one ion is used
to encode the spin while the other ion provides sympathetic cooling of the shared modes of motion. In order to avoid that the cooling lasers couple to the spin transition we choose to work with mixed species ion crystals.
Alternatively, one could rely on single site addressing. The internal levels of the spin ion are described by the Hamiltonian $H_{\rm s} = \hbar\frac{\omega_0}{2}\sigma^z$ while the internal levels of the coolant ion are adiabatically
eliminated from the dynamics leading to the effective description in~\eq{lindblad_dissipator_main} of the cooling~\cite{cirac_laser_cooling, morigi_eit_cooling}.

For concreteness we consider a crystal composed of $^{24}{\rm Mg}^+$ and $^{25}{\rm Mg}^+$. $^{25}{\rm Mg}^+$ has a nuclear spin and we can use the states $\ket{F= 3, m_F = 3} \equiv \ket{\downarrow} $ and  
$\ket{F= 2, m_F = 2} \equiv \ket{\uparrow} $ of the $^2 S_{1/2}$ electronic hyperfine ground-state manifold to encode the spin. The spin can be driven by a microwave or in a two-photon stimulated Raman configuration 
while the desired coupling of the spin to the motional degrees of freedom in the $\sigma^z$ basis is provided by a
``walking standing wave''. In this configuration the spin states are off-resonantly coupled to the $P$ manifold by two laser beams near 280$\,$nm whose beat note is tuned close to one of the motional mode 
frequencies~\cite{quantum_magnet}. The interaction of the ion with the applied fields is described by~\cite{supp_mat}
\beq
H_{\rm int} = \hbar\frac{\Omega_{\rm d}}{2} \sigma^+ \ee^{-\ii \omega_{\rm d} t} 
+ \hbar\frac{\Omega_{\rm odf}}{2} \ee^{ \ii ({\bf k}_{\rm L} {\bf r} + \phi_{\rm L})} \ee^{- \ii \omega_{\rm L} t} \sigma^z + \rm H.c. 
\label{laser_ion_interaction}
\eeq
where $\Omega_{\rm d}$ is the Rabi frequency of the applied microwave or stimulated Raman field and $\omega_{\rm d} \approx \omega_0$ its frequency. $\Omega_{\rm odf}$, ${\bf k}_{\rm L}$, $\omega_{\rm L}$, $\phi_{\rm L}$ 
are the effective laser Rabi frequency, wave vector, frequency and phase, respectively.
Directing ${\bf k}_{\rm L}$ along $z$ the laser only couples to the motion along this axis. A two-ion crystal features two axial modes, an in- and an out-of-phase mode of motion with 
frequencies $\omega_{1,z} \equiv \omega_{1}$ and $\omega_{2,z} \equiv \omega_{2}$. The two modes are well separated in frequency such that choosing the laser frequency $\omega_{\rm L} \approx \omega_{2}$ the spin only couples
to the ouf-of-phase mode. In an interaction picture rotating with the microwave and motional frequencies and under the rotating wave approximation the system's Hamiltonian reads~\cite{supp_mat}
\beq
H_{\rm sb1} = \frac{\hbar\delta}{2} \sigma^z + \frac{\hbar\Omega_{\rm d}}{2} \sigma^x - \frac{\hbar\lambda}{2} (a_{2} + a_{2}^{\dagger} )\sigma^z + \hbar\delta_{\rm m} a_{2}^{\dagger} a_2
\label{sb_ham_ions_main}
\eeq
where  $\delta = \omega_0 - \omega_{\rm d}$ is the detuning of the field driving the spin transition and $\delta_{\rm m} = \omega_{2} - \omega_{\rm L} \ll \omega_{2}$ the detuning of the laser from the motional mode.
The spin-motion coupling is given by $\lambda =-\ii \eta_2 \Omega_{\rm odf} \ee^{\ii( |{\bf k}_{{\rm L}}| z_2^0 + \phi_{\rm L})}$ 
with the Lamb-Dicke factor $\eta_2 = \sqrt{\hbar/(2 m_2 \omega_2)} \tilde{M}_{22} |{\bf k}_{\rm L}|$. Note that the laser phase can be chosen such that $\lambda$ is real. 
$\tilde{M}_{22}$ is the out-of-phase mode amplitude at the spin ion in mass weighted coordinates and $m_2$ its mass. Identifying 
$\hbar \delta = \epsilon$, $\Omega_{\rm d} = -\Delta$ and $\delta_{\rm m} = \omega_{\rm m}$
we obtain the spin-boson Hamiltonian of~\eq{sb_ham_main} for a single mode. Adding the cooling on the second ion the full system evolves according to~\eq{lindblad_main}
where $H=H_{\rm sb1}$ from~\eq{sb_ham_ions_main}.

We simulate the dynamics of the system for experimentally realistic parameters. We consider an axial potential where a single $^{24}{\rm Mg}^+$ ion has a center-of-mass frequency $\omega_{\rm com}/2\pi = 2.54\,$MHz
which leads to an out-of-phase mode frequency $\omega_2/2\pi = 4.36\,$MHz and $\eta_2 \approx 0.15$ for the mixed crystal where we assumed that the lasers inducing the spin-dependent force are at right angles. 
Furthermore, we assume that EIT cooling~\cite{morigi_eit_cooling} is applied to the $^{24}{\rm Mg}^+$ ion which has already been used to sympathetically cool mixed-species ion crystals~\cite{eit_cooling_nist}.
We assume a cooling rate $2\kappa/2\pi = 2.5\,$kHz and a steady-state population $\bar{n}=0.025$ of the mode which is realistic in light of the
results in~\cite{eit_cooling_nist}. Note that one has to make sure that the correspondence to the macroscopic environment holds for the effective mode frequency $\omega_{\rm m} = \delta_{\rm m}$
which is the detuning of the spin-motion coupling and thus much smaller than the physical mode frequency. We chose the field driving the spin to be resonant, i.e. $\epsilon = 0$, and 
a detuning $\omega_{\rm m}/2\pi = 100\,$kHz of the spin-motion coupling such that we recover the parameters we used previously and the correspondence holds. In the simulations we truncate the motional Hilbert 
space at $n_{\rm max} = 15$ excitations which makes truncation errors negligible.

In Fig.~\ref{sb_dynamics} we show the dynamics of $\langle \sigma_z (t) \rangle$ under~\eq{lindblad_main} (where $H=H_{\rm sb1}$ with the parameters from the previous paragraph) for an initial state 
$\rho_0 = \ketbra{\uparrow} \otimes \rho_{\beta}$ where the thermal state $\rho_{\beta}$ has a mean occupation number $\bar{n}=0.025$. We vary the spin-motion coupling $\lambda/2\pi = 10-200\,$kHz.
In panel {\bf a)} we show the dynamics for $\Delta/2\pi = 3\,$kHz. In this case the spin samples the low frequencies of the spectral density in~\eq{jeff_main}. Expanding the spectral density for
small $\omega$ we obtain Ohmic behavior $J_{\rm eff}(\omega) \sim \omega$. We observe a transition from damped to overdamped oscillations with increasing spin-mode coupling $\lambda$. 
This behavior is expected for an Ohmic spectral density at finite temperatures~\cite{weiss_book, leggett_review}. Note, however, that our spectral density $J_{\rm eff} (\omega)$, even if Ohmic for small 
frequencies, does not yield the same correlation function as a strict Ohmic environment. Therefore we can only expect qualitatively similar dynamics~\cite{fkw_lorentzian_sd, mth_lorentzian_sd}.
In panel {\bf b)} we show $\langle \sigma_z (t) \rangle$ for the same initial conditions and $\Delta/2\pi = 100\,{\rm kHz}$ such that the spin is resonant with the mode. 
In this regime the spin dynamics shows a very complex behavior which one would intuitively call non-Markovian. 

{\it Quantification of the degree of non-Markovianity of the dynamics.--} In order to assess the non-Markovian character of the dynamics we compute two measures
of non-Markovianity. The first measure, $\mathcal{N}_{\rm RHP}$, arises from defining Markovianity in terms of the divisibility of the dynamical map of the
dynamics~\cite{rhp_review}, while the second, $\mathcal{N}_{\rm BLP}$, is based on the definition that a dynamics is Markovian when yielding a monotonic decrease of state
distinguishability~\cite{blp_review}.
We evaluated $\mathcal{N}_{\rm RHP}$ numerically~\cite{supp_mat} for the parameters of parts {\bf a)} and {\bf b)} of Fig.~\ref{sb_dynamics} over a time interval $[0,0.01/\Delta]$ and
$[0,0.1/\Delta]$, respectively. The results are shown in part {\bf c)} of the figure. In both cases the measure is non-zero for all couplings $\lambda/2\pi >0 $. 
An evaluation of $\mathcal{N}_{\rm RHP}$ requires process tomography and is therefore experimentally time-consuming already for a single spin. Hence, it might be easier to experimentally
detect non-Markovian dynamics using $\mathcal{N}_{\rm BLP}$ which only requires state tomography.
We numerically computed a lower bound on $\mathcal{N}_{\rm BLP}$~\cite{supp_mat} for the parameters in parts {\bf a)} and {\bf b)} of Fig.~\ref{sb_dynamics} for the whole interval $[0,20/\Delta]$.
The results are shown in part {\bf d)} of the figure. $\mathcal{N}_{\rm BLP}$ witnesses non-Markovianity in all regions where $\mathcal{N}_{\rm RHP}$ does, too.
The somewhat discontinuous behavior of the curve for the resonant case is due to the finite time interval we are sampling.

In order to tailor more complex spectral densities than in this proof-of-principle experiment, one would need to couple the spin to two or more damped modes with the appropriate couplings and cooling
rates that match the effective spectral density to the desired one. In case several modes are used it could be advantageous to use the transverse modes of motion. Due to the smaller bandwidth of the 
transverse phonon frequencies it is easier to couple to and cool several modes at the same time. It should be borne in mind that the cooling rates should be considerably smaller than the spacing between modes.
Only then the damping of each mode can be described by a dissipator as in~\eq{lindblad_dissipator_main}. In order to fill possibly unwanted gaps in the effective spectral density one could then use the
modes of the second transverse direction of motion and place the effective frequencies of these modes between those of the first direction.

Let us finally note that the model can be extended not only to more complex spectral densities by including more ions and thus modes but also to include more spins. Then trapped ions could be used as
a testbed for the dynamics of exciton transport in complex spectral densities and especially the determination of higher order spectral responses, e.g. 2D electronic spectroscopy, which are 
exceedingly hard to compute numerically even for only a few electronic sites and a structured spectral density~\cite{almeida_j_chem_phys}.

In summary, we have shown that spin-boson models with continuous spectral densities can be simulated using damped oscillators in Lindblad description. This leads to a significant reduction of the technical requirements
for the implementation of this paradigmatic model for decoherence and dissipation employing trapped ions. The joint effect of different damped modes allows one to tailor a variety of spectral densities with rich non-Markovian
features. We showed that it is possible to carry out simulations of non-trivial dynamics making use of just one motional mode, and illustrated the practicality of our approach by simulating an experiment with realistic parameters.

{\it Acknowledgements.--} A. L. and D. T. acknowledge very useful discussions with A. Smirne. This work was supported by an Alexander-von-Humboldt Professorship, the ERC synergy grant BioQ and EU projects EQUAM and QUCHIP.
Computational resources were provided by the bwUniCluster and the bwForCluster JUSTUS.

\vspace{5cm}

\newpage

\section{Supplemental Material to ``Simulating spin-boson models with trapped ions''}

\maketitle

\tableofcontents

\appendix

\section{Effective spectral densities of damped harmonic oscillators}

We start by briefly surveying the quantities that we need for the discussion of the effective spectral densities of damped harmonic oscillators.  
We consider the spin-boson model where a spin is coupled to a bath of harmonic oscillators. The spin constitutes
the principal system and the bath consists of an infinite set of independent harmonic oscillators. This is an archetypical model for a two-state system coupled to a dissipative environment 
and is conveniently modeled by the Hamiltonian~\cite{weiss_book_sm}
\beq
H_{\rm sb}=\frac{\epsilon}{2}\sigma^z - \frac{\hbar \Delta}{2}\sigma^x  + \frac{1}{2} \sum_{n} \left[ \frac{p_n^2}{m_n} + m_n \omega_n^2 x_n^2 - c_n q_0 \sigma^z x_n \right]
\label{sb_hamiltonian_sm1}
\eeq
where $\sigma^z = \ketbra{\uparrow} - \ketbra{\downarrow}$ and $\sigma^x = \ketbradif{\uparrow}{\downarrow} + \ketbradif{\downarrow}{\uparrow}$ denote the usual Pauli matrices,
$\epsilon$ the energy splitting of the spin states and $\hbar \Delta$ their coupling. $p_n$ and $x_n$ denote the canonical momenta and coordinates of the environmental modes
of frequency $\omega_n$, $q_0$ is some characteristic length scale and $c_n$ describes the coupling of mode $n$ to the spin. Quantizing the environmental oscillators 
$x_n = \sqrt{\hbar/(2 m_n \omega_n)} (a_n + a_n^{\dagger}) $  so that $a_n$ and $a_n^{\dagger}$ denote the ladder operators of oscillator $n$ we can write the spin-mode coupling as 
\beq
\hbar \lambda_n = c_n q_0 \sqrt{\hbar/(2 m_n \omega_n)} .
\label{coupling_eq}
\eeq
The spin-boson Hamiltonian can then be written as
\beq
H_{\rm sb}=\frac{\epsilon}{2}\sigma^z - \frac{\hbar \Delta}{2}\sigma^x - \frac{1}{2}\sigma^z \sum_{n} \hbar \lambda_{n} (a^{\dagger}_{n} + a_{n}) + \sum_{n} \hbar\omega_{n} a^{\dagger}_{n} a_{n}
\label{sb_hamiltonian_sm2}
\eeq
which is~\eq{sb_ham_main} of the main text. Note that we have omitted the ground-state energies of the oscillators. For an initial product state of spin and environment where the environment
is in a thermal state at inverse temperature $\beta$ the influence of the oscillator environment on the spin is given by the influence functional $F[q,q']$ in~\eq{fv_influence_phase} of the main text
which is in turn determined by the reservoir correlation function~\cite{weiss_book_sm} 
\beq
L(t) = \frac{1}{\hbar^2} \langle X(t) X(0) \rangle_{\beta}
\label{coordinate_correlation_first}
\eeq
with the collective coordinate $X(t) = q_0 \sum_n c_n x_n = \sum_n \hbar \lambda_n (a_n +a_n^{\dagger})$. 
The reservoir correlation function can be equivalently given in terms of the spectral density $J(\omega)$ 
\beq
L(t) = \frac{1}{\pi} \int_0^{\infty} {\rm d}\omega \,J(\omega)  \left[ \coth \left( \frac{\beta \hbar \omega}{2} \right) \cos( \omega t) - \ii \sin( \omega t)\right].
\label{lt_sd_sm}
\eeq
It is known that an oscillator damped by a bath with Ohmic spectral density produces an effective environment with Lorentzian spectral density~\cite{garg_lorentzian_sd_sm}.
Here, we inspect in more detail when the same can be done for the damped harmonic oscillator in Lindblad description. To this end, it is instructive to start from the time domain
and consider $L(t)$ for the two cases.

\subsection{Time domain considerations}

The reservoir correlation function $L(t)$ in~\eq{coordinate_correlation_first} may be written explicitly in terms of the environmental coordinate correlation functions using
$X(t) = q_0 \sum_n c_n x_n $
\beq
L(t) = q_0^2 \sum_n \frac{c_n^2}{\hbar^2} \langle x_n(t) x_n(0) \rangle_{\beta}
\label{coordinate_correlation}
\eeq
where we have used that the oscillators are independent.
In the following we consider only a single oscillator and therefore omit the index $n$ from now on. The function $\langle x(t) x(0) \rangle_{\beta}$ is in general a complex function
and we can write it in terms of its real and imaginary parts
\beq
\langle x(t) x(0) \rangle_{\beta} = S(t) + \ii A(t)
\eeq
where
\begin{eqnarray}
S(t) &= \frac{1}{2} \langle \{ x(t),x(0) \} \rangle_{\rm \beta},\\
A(t) &= \frac{1}{2\ii} \langle [ x(t),x(0) ] \rangle_{\rm \beta}.
\end{eqnarray}
The imaginary part $A(t)$ is related to the damped oscillator's response function $\chi(t)$ through $\chi(t)= -\frac{2}{\hbar} \Theta (t) A(t)$~\cite{weiss_book_sm}
where $\Theta (t)$ is the Heaviside step function. Note that accordingly also $L(t)$ is a complex function
\beq
L(t) = L'(t) + \ii L''(t).
\eeq
Let us now consider a damped oscillator that evolves according to the Lindblad equation given in~\eq{lindblad_main} of the main text
\beq
\dot{\rho} = -\ii[\omega_{\rm m} a^{\dagger} a, \rho] + \mathcal{D}_{\kappa,\bar{n}} \rho
\label{lindblad_damped_ho_app}
\eeq
with dissipator
\beq
\begin{split}
\mathcal{D}_{\kappa,\bar{n}} \rho = &\kappa (\bar{n} +1) [a \rho  a^{\dagger} - a^{\dagger} a \rho ] \\
& + \kappa \bar{n} [a^{\dagger} \rho a - a a^{\dagger} \rho ] + {\rm H.c.}
\end{split}
\label{lindblad_dissipator}
\eeq
given in~\eq{lindblad_dissipator_main} of the main text.
The above dissipator takes the mode populations to a thermal state with mean occupation number $\bar{n}$ at a rate $2\kappa$.
We can compute the coordinate correlation function $\langle x(t) x(0) \rangle_{\beta,{\rm L}} = S_{\rm L}(t) + \ii A_{\rm L}(t)$ of the damped harmonic oscillator
in Lindblad description using the quantum regression theorem:
\beq
S_{\rm L} (t) = \frac{\hbar}{2 m \omega_{\rm m}} \coth\left(\frac{\beta \hbar \omega_{\rm m}}{2}\right) \,\cos(\omega_{\rm m} t)  \ee^{-\kappa |t|} 
\label{st_lindblad_app}
\eeq
and
\beq
A_{\rm L} (t) = -\frac{\hbar}{2 m \omega_{\rm m}} \sin(\omega_{\rm m} t) \ee^{-\kappa |t|}.
\label{at_lindblad_app}
\eeq
Here $m$ is the mass of the oscillator. Note that the frequency $\omega_{\rm m}$ is taken to include possible renormalizations of the mode frequency due to the damping
and $\kappa \ll \omega_{\rm m}$ is necessary to derive the Lindblad equation above.
Inserting the result into~\eq{coordinate_correlation} and using~\eq{coupling_eq} we obtain the real and imaginary parts $L_{\rm L}' (t)$ and $ L_{\rm L}'' (t)$ 
of $L_{\rm L} (t)$ from Eqs.~\eqref{l1l_main} and~\eqref{l_imag} of the main text
\beq
L'_{\rm L} (t) = \lambda^2 \coth\left(\frac{\beta \hbar \omega_{\rm m}}{2}\right) \,\cos(\omega_{\rm m} t)  \ee^{-\kappa |t|} 
\label{l_real_lindblad_app}
\eeq
and
\beq
L''_{\rm L} (t) = -\lambda^2 \sin(\omega_{\rm m} t) \ee^{-\kappa |t|}.
\label{l_imag_lindblad_app}
\eeq
As we stated earlier we have  $\chi(t)= -\frac{2}{\hbar} \Theta (t) A(t)$. Inserting $A_{\rm L} (t)$ into the previous equation yields the response function of the classical 
damped harmonic oscillator. Having in mind that an Ohmic spectral density leads to the classical equation of motion for a damped oscillator~\cite{weiss_book_sm}, and thus the same 
response function, it seems appropriate to compare the regression theorem results to that of the oscillator damped by an Ohmic bath.

Therefore, we move on to the harmonic oscillator damped by a thermal oscillator bath with Ohmic spectral density. For this case, it is also possible to calculate the coordinate correlation 
function $\langle x(t) x(0) \rangle_{\beta}$ analytically~\cite{grabert_damped_qho_sm, weiss_book_sm}. We denote the free oscillation frequency of the oscillator by $\Omega$ while we denote 
the damping rate on the oscillator's coordinate by $\kappa_{\rm ohm}$. In the underdamped regime $\kappa_{\rm ohm} < \Omega$ the oscillator's frequency is reduced to
$\omega_{\rm r} = \sqrt{\Omega^2 - \kappa_{\rm ohm}^2}$ due to the damping. Since we want to compare the results to the Lindblad case where $\kappa \ll \omega_{\rm m}$ we will always
have $\kappa_{\rm ohm} \ll \Omega$ such that we are in the underdamped regime. In this regime the real part of the coordinate correlation $S(t)$ splits in two parts~\cite{weiss_book_sm}
\beq
S (t) = S_1(t) + S_2(t)
\label{def_st_app}
\eeq
with
\beq
\begin{split}
S_1(t) =& \frac{\hbar}{2 m \omega_{\rm r}} \left[\frac{\sinh(\beta \hbar \omega_{\rm r})}{\cosh(\beta \hbar \omega_{\rm r}) -\cos( \beta \hbar \kappa_{\rm ohm})} \cos(\omega_{\rm r} t) \right. \\
&\left.  +\frac{\sin(\beta \hbar \kappa_{\rm ohm})}{\cosh(\beta \hbar \omega_{\rm r}) -\cos(\beta \hbar \kappa_{\rm ohm})} \sin(\omega_{\rm r} |t|) \right] \ee^{-\kappa_{\rm ohm} |t|} \\
\end{split}
\label{def_s1_app}
\eeq
and
\beq
S_2(t) = -\frac{4\kappa_{\rm ohm}}{m\beta} \sum_{n=1}^{\infty} \frac{\nu_n \ee^{-\nu_n |t|}}{(\Omega^2 + \nu_n^2)^2 - 4\kappa_{\rm ohm}^2 \nu_n^2}
\label{def_s2_app}
\eeq
where the $\nu_n= 2\pi n/(\hbar \beta)$ are the Matsubara frequencies. The imaginary part reads
\beq
A(t) = -\frac{\hbar}{2 m \omega_{\rm r}} \sin(\omega_{\rm r} t) \ee^{-\kappa_{\rm ohm} |t|}.
\label{at_ohmic_app}
\eeq
Comparing Eqs.~\eqref{at_ohmic_app} and~\eqref{at_lindblad_app} we see that the imaginary parts $A(t)$ and $A_{\rm L} (t)$ are exactly equal for $\omega_{\rm r} = \omega_{\rm m}$ and
$\kappa_{\rm ohm} = \kappa$ which we will assume from now on. With this substitution and inserting Eqs.~\eqref{def_s1_app}-\eqref{at_ohmic_app} into~\eq{coordinate_correlation} we obtain
$L(t) = L' (t) +\ii L''(t)=L_1(t)+L_2(t)+\ii L''(t)$ where
\beq
\begin{split}
L_1(t) &= \lambda^2 \left[\frac{\sinh(\beta \hbar \omega_{\rm m})}{\cosh(\beta \hbar \omega_{\rm m}) -\cos(\beta \hbar \kappa)} \cos(\omega_{\rm m} t) \right.  \\
& \left. +\frac{\sin(\beta \hbar \kappa)}{\cosh(\beta \hbar \omega_{\rm m}) -\cos(\beta \hbar \kappa)} \sin(\omega_{\rm m} |t|) \right]\ee^{-\kappa |t|},
 \\ 
L_2(t) &= -\lambda^2 \frac{8 \kappa \omega_{\rm m}}{\hbar \beta} \sum_{n=1}^{\infty} \frac{\nu_n \ee^{-\nu_n t}}{(\Omega^2 + \nu_n^2)^2 - 4\kappa^2 \nu_n^2} \\
\end{split}
\label{l_real_app}
\eeq
and
\beq
L'' (t) = -\lambda^2 \sin(\omega_{\rm m} t) \ee^{-\kappa |t|}
\label{l_imag_app}
\eeq
recovering Eqs.~\eqref{l_real} and~\eq{l_imag} of the main text.

The symmetric parts $S(t)$ and $S_{\rm L} (t)$ do not coincide after the substitution $\omega_{\rm r} = \omega_{\rm m}$ and $\kappa_{\rm ohm} = \kappa$. Hence, in the following we seek the regimes where the two functions coincide.
In order to identify $S_{\rm L} (t)$ with $S (t)$ we need to be able to neglect $S_2 (t)$ as well as the sine component in $S_1 (t)$. We start by considering $S_2 (t)$. The argument follows
Refs.~\cite{grabert_damped_qho_sm, talkner_fdt_sm}. The  Matsubara frequencies $\nu_n$ determine the time scale on which $S_2 (t)$ decays, the smallest decay rate being $\nu_1$. Accordingly, if the
decay rate $\kappa$ is much smaller than the smallest Matsubara frequency, $S_2(t)$ drops to zero much faster than $S_1(t)$. This is the regime where
\beq
\frac{\kappa \hbar \beta }{2\pi} = \frac{\kappa}{\nu_1} \ll 1.
\label{lower_t_limit}
\eeq
In this regime one expects that $S_2(t)$ will only produce deviations on very short time scales and is negligible if we are interested in not too short time scales. This is the case in our
considerations. If $S_2 (0) \ll S_1 (0)$ we can neglect $S_2(t)$ completely.

Assuming we can disregard $S_2(t)$ we need to find the regime where 
\beq
S_{\rm L}(t) \approx S_1(t).
\eeq
In the limit $\beta \hbar \kappa \ll 1$ we can expand the sine and cosine terms in $S_1(t)$ in this small parameter and to first order we obtain
\beq
\begin{split}
S_1(t) \approx & \frac{\hbar}{2 m \omega_{\rm m}} \left[\frac{\sinh(\beta \hbar \omega_{\rm m})}{\cosh(\beta \hbar \omega_{\rm m}) -1 } \cos(\omega_{\rm m} t) \right. \\
& \left. +\frac{\hbar \beta \kappa}{\cosh(\beta \hbar \omega_{\rm m}) -1 } \sin(\omega_{\rm m} |t|) \right]\ee^{-\kappa |t|} \\
\approx & \frac{\hbar}{2 m \omega_{\rm m}} \frac{\sinh(\beta \hbar \omega_{\rm m})}{\cosh(\beta \hbar \omega_{\rm m}) -1} \cos(\omega_{\rm m} t) \ee^{-\kappa |t|}
\end{split}
\eeq
where we have used $\hbar \beta \kappa \ll \sinh(\hbar \beta \omega_{\rm m})$ in the last step. Using the identity $\coth \frac{x}{2} = \sinh(x)/(\cosh x -1)$ finally yields $S_1(t) = S_{\rm L}(t)$
if the reservoirs are at the same inverse temperature $\beta$. Accordingly, we assume that the reservoir in the Lindblad description and the Ohmic oscillator bath have the same inverse temperature $\beta$
from now on.

Thus, we have established a regime where the coordinate correlation function of the Lindblad damped harmonic oscillator approximately coincides with that of an oscillator damped by a reservoir with
Ohmic spectral density. In this regime the Lindblad damped oscillator should act as a macroscopic reservoir with Lorentzian spectral density as in~\eq{jeff_main} of the main text.

Note that for a given cooling rate $\kappa$ the condition in~\eq{lower_t_limit} puts a lower bound on the temperature where we can neglect $S_2(t)$ and thus a lower bound on the temperature 
where the Lindblad damped oscillator produces the same coordinate correlation function as the oscillator damped by a reservoir with Ohmic spectral density. Thus, we require $\kappa \ll \omega_{\rm m}$
and $\frac{\kappa \hbar \beta }{2\pi} \ll 1$ to make the identification. Indeed, Refs.~\cite{grabert_damped_qho_sm, talkner_fdt_sm} estimate that the quantum regression theorem can only yield
quantitatively correct predictions for the two-time correlation functions of the damped harmonic oscillator if the two above conditions are met.

For ion-trap experiments one usually considers the mean occupation number $\bar{n}$ of the bosonic modes rather than their temperature and therefore it is desirable
to cast condition~\eqref{lower_t_limit} in a form where it depends on $\bar{n}$. Assuming a thermal state for a bosonic mode we can associate the temperature
$T_{\rm eff} = \hbar \omega/[k_B \log(1+1/\bar{n})]$ to the mode and the condition in~\eq{lower_t_limit} becomes
\beq
\frac{\log (1+\frac{1}{\bar{n}})}{2\pi} \frac{\kappa}{\omega_{\rm m}} \ll 1.
\eeq
Note that in the ion-trap implementation the mode frequency is an effective frequency much smaller than the physical frequency of the mode (see App. \ref{sec: spin-boson with ions}). 
Therefore, one has to make sure the above condition is met for the effective frequency such that the correspondence to the effective harmonic environment is not lost.

\begin{figure}[hbt]
\includegraphics[width=.9\columnwidth]{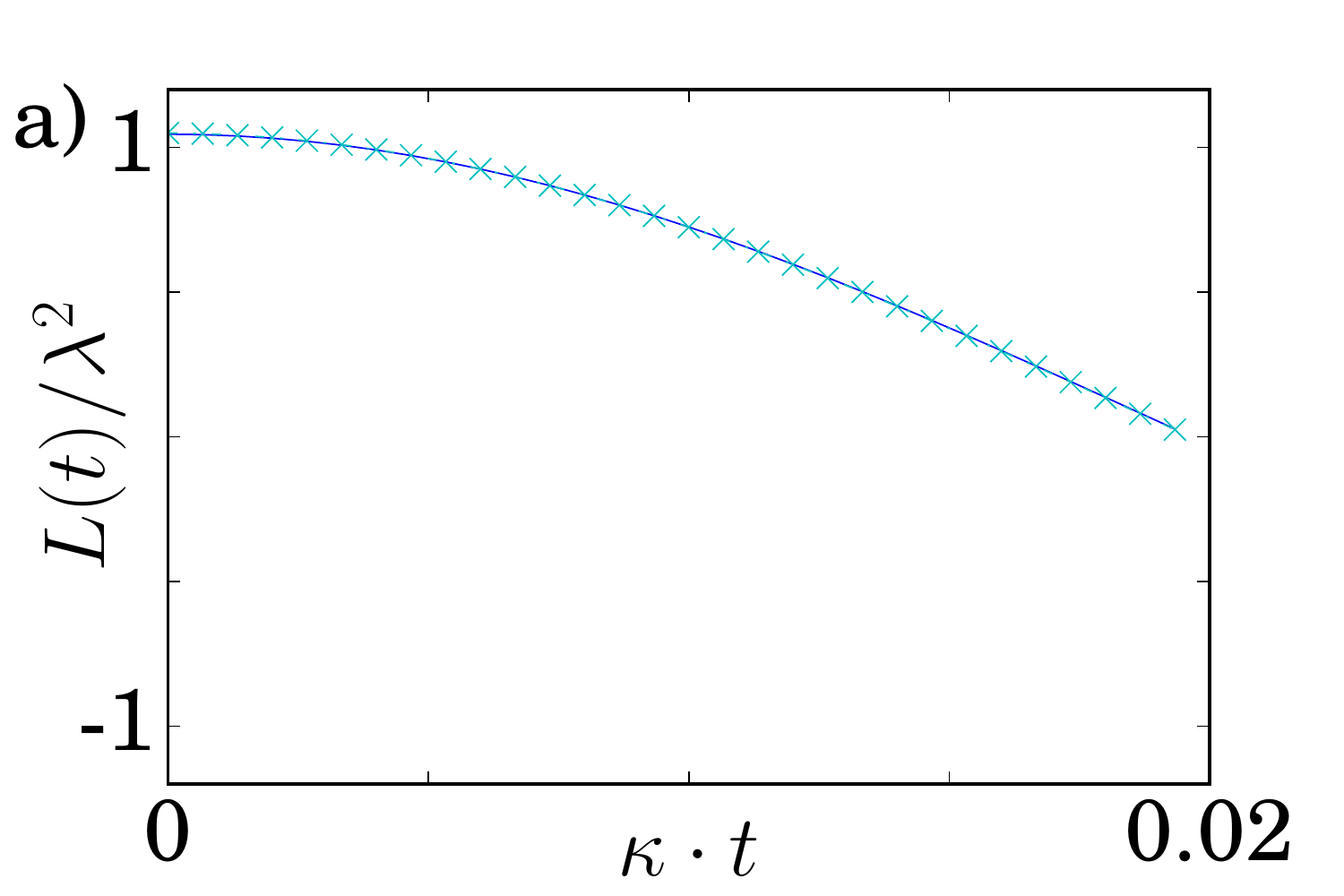}
\includegraphics[width=.9\columnwidth]{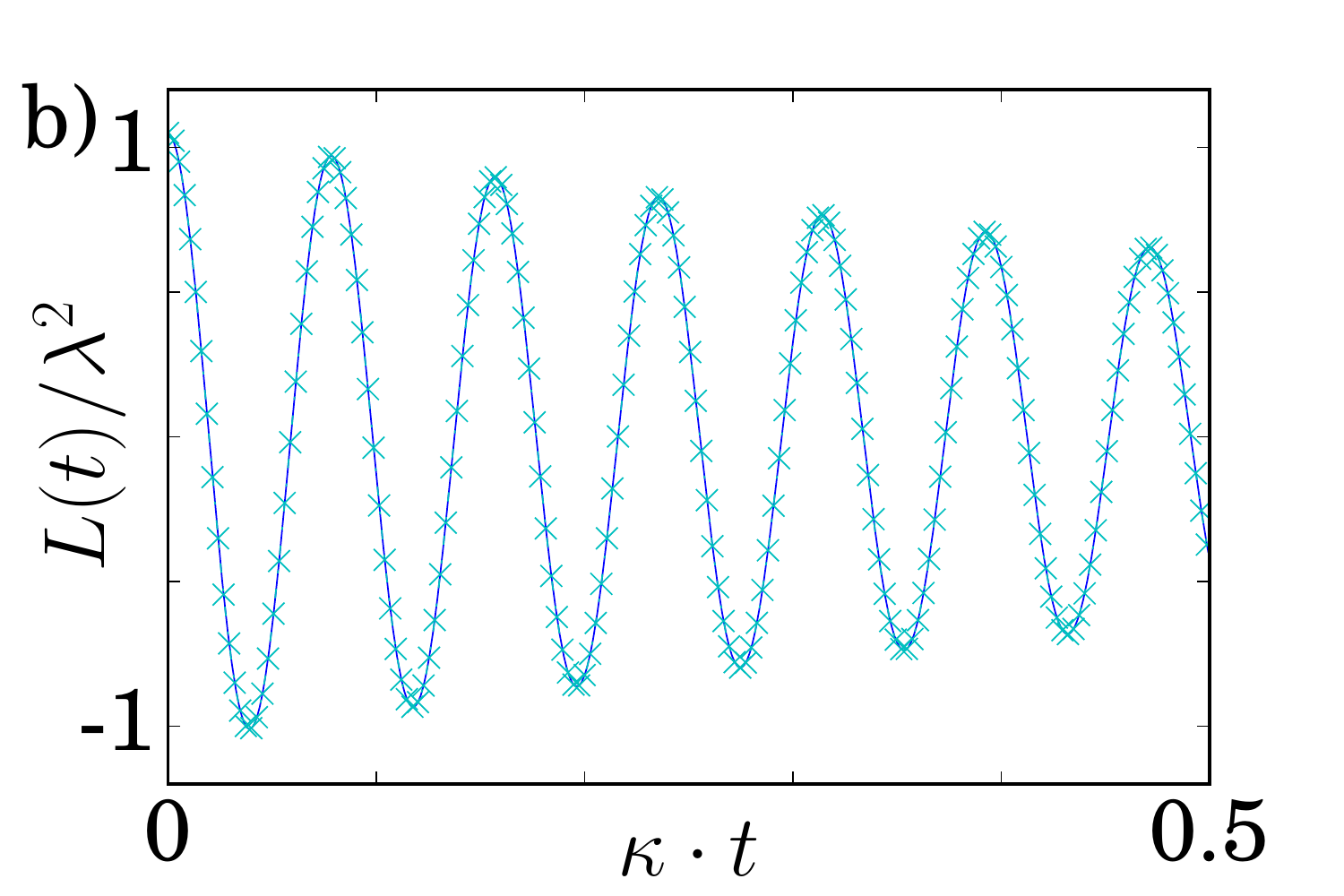}
\caption{Comparison of $L'(t) = L_1(t)+L_2(t)$ from~\eq{l_real_app} including the first $10^4$ Matsubara frequencies (blue solid lines) and $L_{\rm L}'(t)$ from~\eq{l_real_lindblad_app} (dashed-dot line and crosses) for
$\omega_m/2\pi = 100\,$kHz, $\kappa/2\pi=1.25\,{\rm kHz}$ and $\bar{n}(\omega_{\rm m})=0.025$ ($\hbar \beta = 5.91 \cdot 10^{-6}\,$s). Panel {\bf a)} shows the time evolution for short times, while panel {\bf b)} illustrates the intermediate time behavior.}
\label{corr_fun_match}
\end{figure}

In order to make the above considerations more quantitative and illustrate that the match of the reservoir correlation functions is indeed very good we make a numerical comparison of the functions $L(t)$ and $L_{\rm L} (t)$
in the regime $\kappa \ll \omega_{\rm m}, \nu_1$. Since the imaginary parts of the two functions are equal we focus on the real parts $L'(t)$ and $L_{\rm L} '(t)$.
In Fig.~\ref{corr_fun_match} we plot $L'(t)/\lambda^2$ including the first $10^4$ Matsubara frequencies together with $L_{\rm L} (t)$ for $\omega_{\rm m}/2\pi = 100\,$kHz, $\kappa/2\pi=1.25\,$kHz and a mean occupation
number $\bar{n} (\omega_{\rm m})=0.025$ which corresponds to $\hbar \beta = 5.91 \cdot 10^{-6}\,$s. These parameters are realistic in an ion trap experiment. In part {\bf a)} of the figure we compare $L'(t)$ and $L_{\rm L}' (t)$
on short and in part {\bf b)} on intermediate time scales. One can appreciate excellent agreement between the two functions.

\begin{figure}[hbt]
\includegraphics[width=.9\columnwidth]{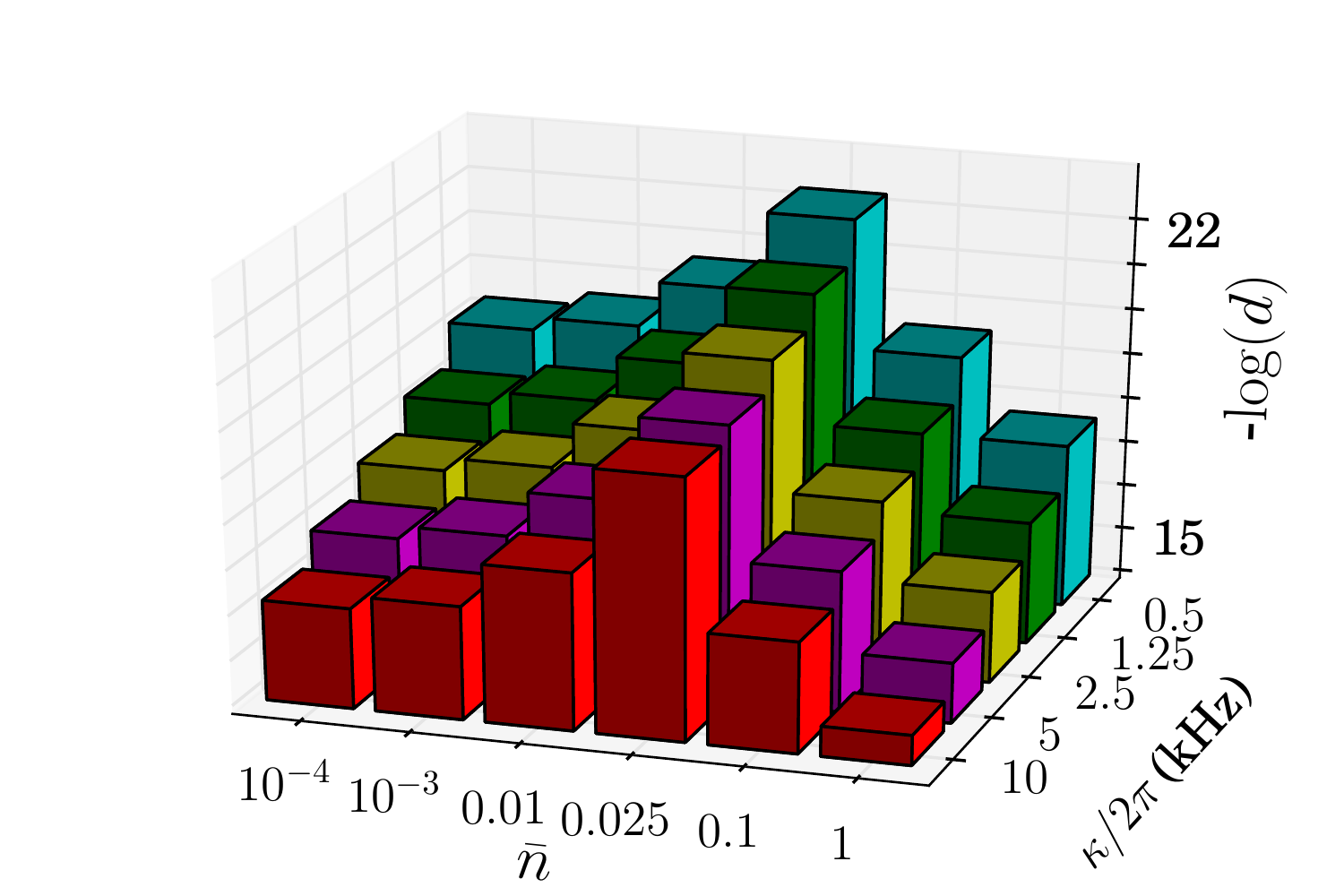}
\caption{ The figure shows the distance $d$,~\eq{def_d}, between the correlation functions $L(t)$ including the first $10^4$ Matsubara frequencies and $L_{\rm L} (t)$ for different values of the cooling 
rate $\kappa$ and mean occupation number $\bar{n}$ for fixed mode frequency $\omega_{\rm m}/2\pi=100\,$kHz. Higher bars correspond to smaller values of $d$.}
\label{fig_deviations}
\end{figure}

In order to illustrate for which parameters the approximation works well we compute the distance
\beq
d =  \frac{1}{\lambda^2} \left| \int_0^{\infty} {\rm d}t [L(t)-L_{\rm L}(t)]\,\right| 
\label{def_d}
\eeq
between the functions $L(t)$ and $L_{\rm L}(t)$ which can be evaluated analytically to yield
\beq
\begin{split}
d = & c_{\rm q} \frac{\kappa}{\kappa^2+\omega_{\rm m}^2} + c_{\rm cl} \frac{\omega_{\rm m}}{\kappa^2+\omega_{\rm m}^2} \\
&- \frac{8 \kappa \omega_{\rm m}}{\hbar \beta} \sum_{n=1}^{\infty} \frac{1}{(\omega_{\rm m}^2 + \kappa^2 + \nu_n^2)^2 - 4\kappa^2 \nu_n^2}, 
\end{split}
\eeq
where we used the abbreviations
\begin{equation*}
\begin{split}
c_{\rm q} &= \frac{\sinh(\beta \hbar \omega_{\rm m})}{\cosh(\beta \hbar \omega_{\rm m}) -\cos(\hbar \beta \kappa)} - \coth\left(\frac{\hbar \beta \omega_{\rm m}}{2}\right), \\
c_{\rm cl} &=  \frac{\sin(\beta \hbar \kappa)}{\cosh(\beta \hbar \omega_{\rm m}) -\cos(\hbar \beta \kappa)}. 
\end{split}  
\end{equation*}
We evaluate the difference for different cooling rates and mean occupation numbers while keeping the mode frequency fixed at $\omega_{\rm m}/2\pi = 100\,$kHz.
The results are depicted in Fig.~\ref{fig_deviations}. Note that higher bars in the figure correspond to smaller values of $d$. We observe that increasing $\kappa$ increases the difference between
the two functions. For a fixed cooling rate we observe that the distance is minimal for intermediate values of $\bar{n}$. 
This can be understood by considering Eqs.~\eqref{l_real_lindblad_app} and~\eqref{l_real_app}. In order to identify $L(t)$ and $L_{\rm L} (t)$ we need to be able to neglect $L_2(t)$ and the sine 
component in $L_1(t)$. The condition in~\eq{lower_t_limit} provides the regime where $L_2(t)$ is negligible and favors higher temperatures.
However, in order to suppress the sine component in $L_1(t)$ lower temperatures are more favorable. Thus, we obtain the best match for intermediate temperatures.

\subsection{Frequency space considerations}

In~\cite{garg_lorentzian_sd_sm} Garg {\it et al.} show that a harmonic oscillator which in turn is damped by an oscillator environment with Ohmic spectral density with infinite cutoff produces the effective spectral density
\beq
\frac{q_0^2}{\hbar} J_{\rm eff, ohm} (\omega) = \frac{q_0^2 }{\hbar} \frac{c_1^2}{m} \frac{2\kappa \omega}{(\Omega^2-\omega^2)^2 + 4 \omega^2 \kappa^2}.
\eeq
Here, $\kappa$ is the damping induced by the bath on the coordinate of the oscillator, $\Omega$ and $m$ its free oscillation frequency and mass, respectively, and $q_0 c_1$ its coupling to the spin [see~\eq{sb_hamiltonian_sm1}]. 
Note that upon writing the influence functional as in~\eq{fv_influence_phase} of the main text we have absorbed the prefactor $q_0^2/\hbar$ into the spectral density. 
Using~\eq{coupling_eq} and setting $\omega_{\rm m}^2 = \Omega^2 - \kappa^2 $ one obtains the spectral density $J_{\rm eff} (\omega)$ in~\eq{jeff_main} of the main text
\beq
J_{\rm eff} (\omega) = \lambda^2 \left[\frac{\kappa}{\kappa^2 + (\omega-\omega_{\rm m})^2} - \frac{\kappa}{\kappa^2 + (\omega+\omega_{\rm m})^2} \right].
\label{eff_sd_garg_sm}
\eeq

In the previous section we have seen that in a certain parameter regime the Lindblad description of the damped harmonic oscillator reproduces the coordinate correlation function and thus $L(t)$ of the oscillator
damped by an Ohmic bath. $L(t)$ can also be written in terms of the spectral density $J(\omega)$ according to~\eq{lt_sd_sm}. In fact, in almost all cases environments are
characterized by their spectral density rather than their correlation functions. Therefore, we analyze the effective spectral density of the Lindblad-damped oscillator and compare it to the 
Lorentzian spectral density $J_{\rm eff} (\omega)$ in~\eq{eff_sd_garg_sm} above.

The Fourier representation of $L_{\rm L} (t)$ in Eqs.~\eqref{l_real_lindblad_app} and~\eqref{l_imag_lindblad_app} reads
\beq
\begin{split}
L_{\rm L}(t) = \frac{1}{\pi} \int_0^{\infty} {\rm d} \omega & \left[  \tilde{J}_{\rm eff} (\omega) \coth \left( \frac{\hbar \beta \omega}{2} \right) \cos(\omega t) \right. \\
&  -\ii \left.  J_{\rm eff} (\omega) \sin(\omega t) \right]
\end{split}
\label{fluctuations_lindblad}
\eeq
with $J_{\rm eff} (\omega)$ as in~\eq{eff_sd_garg_sm} above and 
\beq
\tilde{J}_{\rm eff}(\omega) = \lambda^2 \frac{\coth\left(\frac{\beta\hbar\omega_{\rm m}}{2}\right)}{\coth\left(\frac{\beta\hbar\omega}{2}\right)} \left[\frac{\kappa}{\kappa^2+(\omega-\omega_{\rm m})^2} +\frac{\kappa}{\kappa^2+(\omega+\omega_{\rm m})^2}\right].
\eeq
We note that in general $\tilde{J}_{\rm eff} (\omega) \neq J_{\rm eff} (\omega)$ and hence we cannot write $L_{\rm L} (t)$ as a function of a single spectral density as in~\eq{lt_sd_sm}, in general. 
Yet, from our considerations in the previous section we expect that for appropriate parameters
\beq
\tilde{J}_{\rm eff} (\omega) \approx J_{\rm eff} (\omega)
\label{freq_space_match}
\eeq
such that we obtain the form of $L(t)$ in~\eq{lt_sd_sm} as for a macroscopic environment.

\begin{figure}[hbt]
\centering
\includegraphics[width=.95\columnwidth]{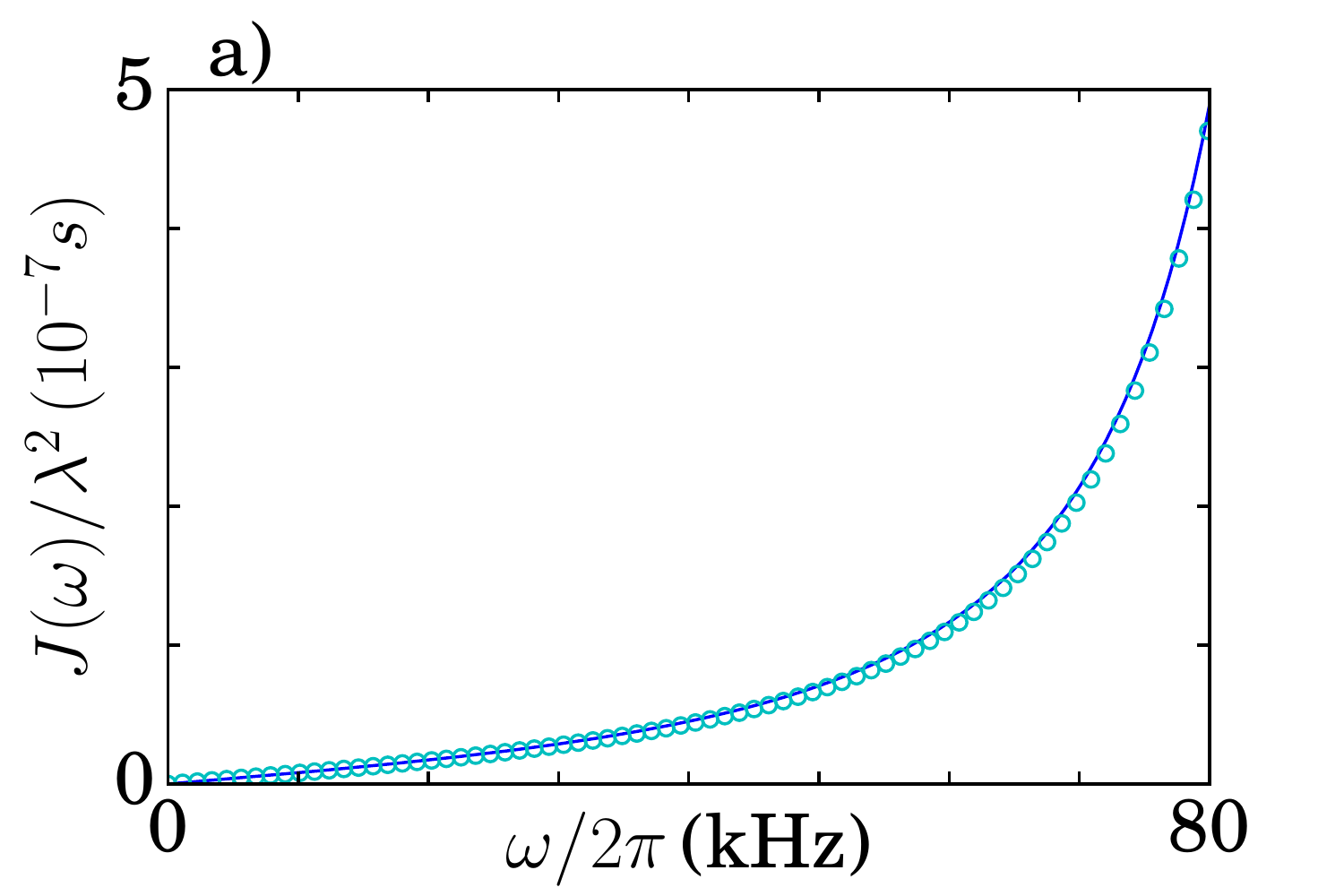}\hfill
\includegraphics[width=.95\columnwidth]{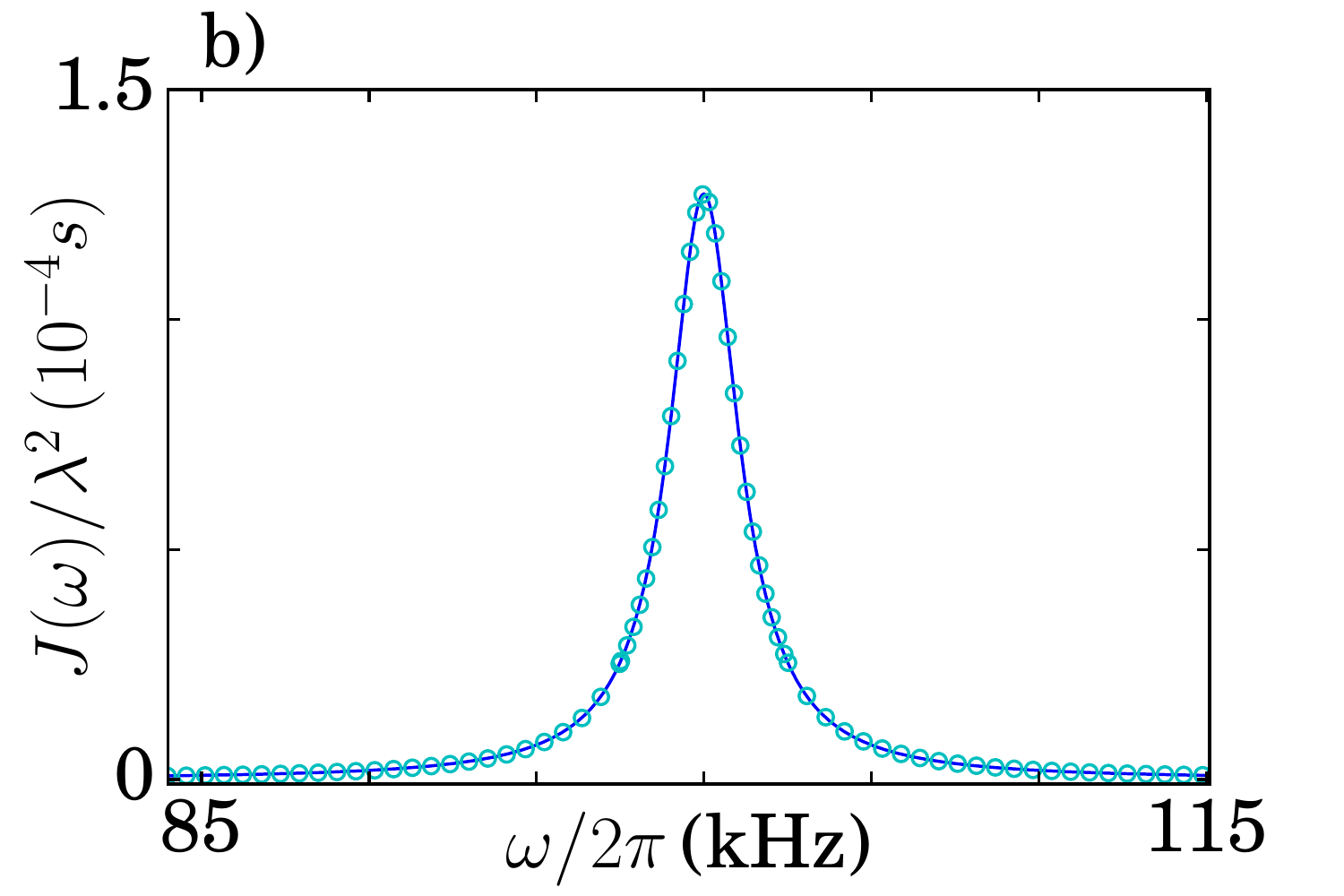}\hfill
\includegraphics[width=.95\columnwidth]{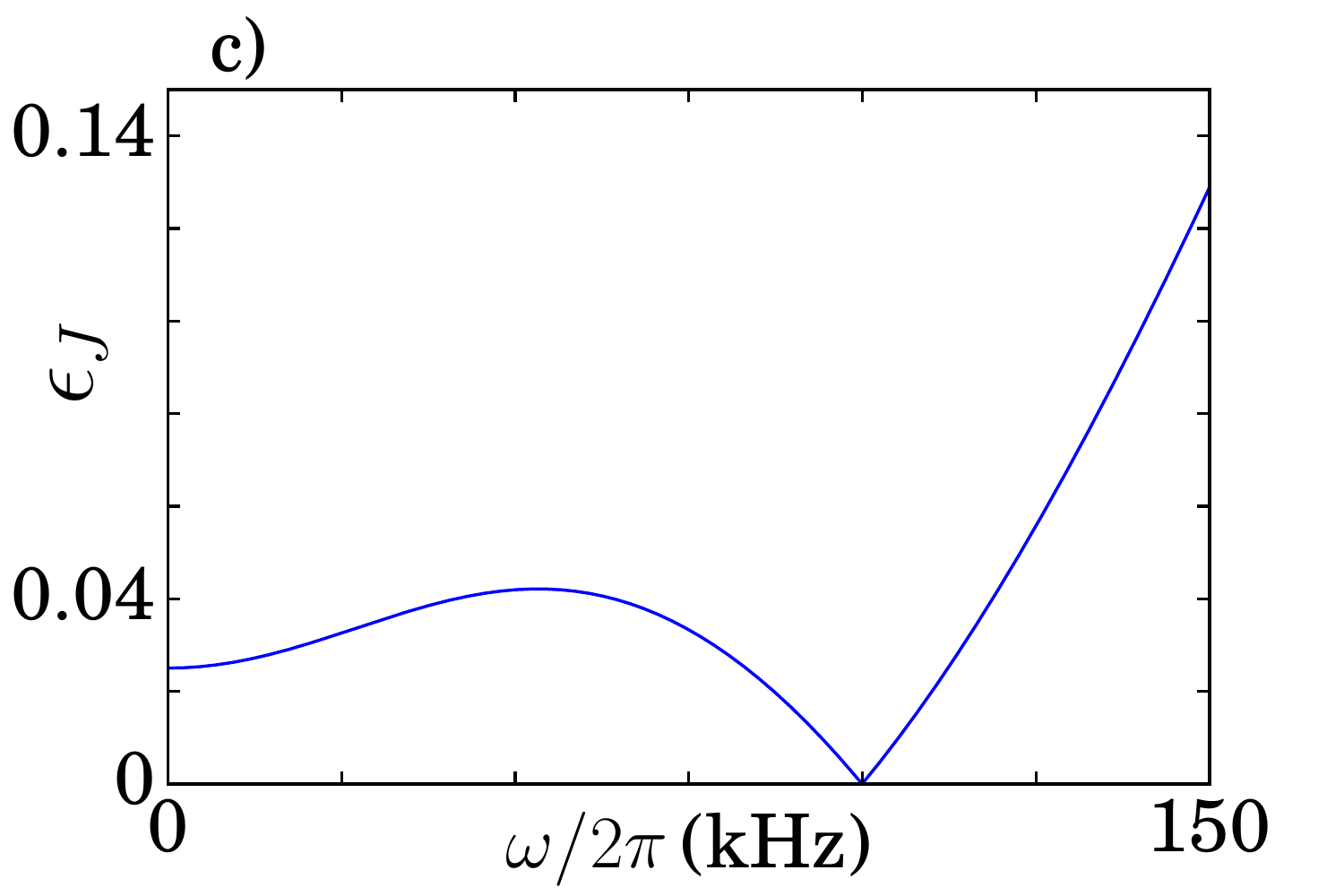}
\caption{The figure compares $J_{\rm eff} (\omega) $ (solid line) and $\tilde{J}_{\rm eff} (\omega)$ (circles) from~\eq{freq_space_match} for $\omega_{\rm m}/2\pi= 100\,$kHz, $\kappa/2\pi = 1.25\,$kHz
and $\bar{n} = 0.025$. Part {\bf a)} shows the behavior for small frequencies while part {\bf b)} depicts the two functions around the resonance $\omega_{\rm m}/2\pi = 100\,$kHz. In part {\bf c)} we
show the relative error $\epsilon_{J}$ from~\eq{def_epsilon} over the relevant frequency range covered by the spectral density.}
\label{fig_freq_space}
\end{figure}

In Fig.~\ref{fig_freq_space} we compare the left and right hand sides of~\eq{freq_space_match} for the parameters we use in the previous section and the main text, i.e.
 $\omega_{\rm m}/2\pi= 100\,$kHz, $\kappa/2\pi = 1.25\,$kHz and  $\bar{n}(\omega_{\rm m})=0.025$ ($\hbar \beta = 5.91 \cdot 10^{-6}\,$s) where we found very good agreement between the correlation functions $L(t)$ and $L_{\rm L} (t)$ 
(see Fig.~\ref{corr_fun_match}). Panel {\bf a)} shows $J_{\rm eff} (\omega)$ (solid line) and $\tilde{J}_{\rm eff} (\omega)$ (circles) for low frequencies and part {\bf b)} shows 
the behavior around the resonance $\omega_{\rm m}/2\pi=100\,$kHz. Both parts of the figure show that we obtain very good agreement in frequency space, too. 
Part {\bf c)} of the figure shows the relative error
\beq
\epsilon_J = \frac{|\tilde{J}_{\rm eff}(\omega)-J_{\rm eff}(\omega)|}{J_{\rm eff}(\omega)}
\label{def_epsilon}
\eeq
which is remarkably small over the whole range $\omega/2\pi = 0-150\,$kHz. Note that the increase in the relative error for higher frequencies is because the spectral density goes to
zero more rapidly than the effective one. However, since both contributions are small the effect of this difference should be negligible.

Thus, we confirm the result of the previous section: for appropriate choices of mode frequency,
cooling rate and temperature, the damped oscillator evolving according to the Lindblad equation can be attributed the effective spectral density $J_{\rm eff} (\omega)$ of a macroscopic
oscillator environment. Note that the treatment is not perturbative in the spin-motion coupling $\lambda$, so that this equivalence is valid for arbitrary values of $\lambda$.

%In some cases it might only be required to model a certain range of a spectral density. In this case one can add a weight function to~\eq{min_procedure} to make the approximation better in this region.
%Note that in any case the optimization is subject to the condition $\kappa_n \ll \omega_n$. 

\section{tDMRG simulations using the TEDOPA algorithm}
For macroscopic environments the  Hamiltonian for the
spin-boson model considered in this work becomes
\begin{align}
	&H = H_{\text{sys}} + H_{\text{env}} + H_{\text{int}},
	\label{eq:H_sb1}\\
    &H_{\text{sys}} = \frac{\epsilon}{2}\sigma^z-\frac{\hbar\Delta}{2} \sigma^x, \label{eq:hsys}\\ 
&H_{\text{env}} = \hbar \int_0^{\omega_{\text{max}}}  {\rm d}\omega\,\omega a_\omega^\dagger a_\omega,
	\label{eq:H_sb2}\\
    & H_{\text{int}} =-\sigma^z \frac{\hbar}{2}   \int_0^{\omega_{\text{max}}}  {\rm d} \omega \,h (\omega)
			  (a_\omega + a_\omega^\dagger),
	\label{eq:H_sb3}
\end{align}
where we have introduced a hard cutoff $\omega_{\text{max}}$.
The spectral density $J(\omega)$ is given by
\begin{equation}
	J(\omega) = \pi h^2(\omega).
	\label{eq:def-sd}
\end{equation}
To simulate the evolution of the spin-boson model, we resorted to the Time Evolving density matrix with orthogonal polynomials (TEDOPA) algorithm.
In this section we briefly present the TEDOPA scheme and refer to \cite{prior_sm, chain_mapping_sm} for
a more detailed presentation of the algorithm. TEDOPA is a certifiable and numerically exact method
to treat open quantum system dynamics \cite{prior_sm, ref:woods2015_sm}.

In a two-stage process TEDOPA first employs a unitary transformation reshaping the spin-boson model into a one-dimensional
configuration. New oscillators with creation and annihilation operators $b_n^\dagger$ and $b_n$ are defined using the unitary transformations $U_n(\omega)$ 
\begin{align}
	&U_n (\omega) = h (\omega) p_n (\omega), \\
	&b_n^\dagger =
	\int_0^{\omega_{\text{max}}} {\rm d}\omega\, U_n (\omega) a_\omega^\dagger,
	\label{eq:chainmap}
\end{align}
where  $p_n (\omega),\ n=0,1,\ldots$ are orthogonal polynomials with
respect to the measure
${\rm d}\mu(\omega)=h^2(\omega){\rm d}\omega$~\cite{chain_mapping_sm}. While in certain
cases it is possible to perform this transformation analytically~\cite{chain_mapping_sm}, in
general a numerically stable procedure is used~\cite{ref:gautschi1994_sm}. This transformation maps the environment to a semi-infinite one-dimensional chain of oscillators with
nearest-neighbor interactions. In this configuration the spin only interacts with the first site of the chain.
The Hamiltonian~\eqref{eq:H_sb1} becomes
\begin{align}
	H =&
	H_{\text{sys}} 
	-\hbar \frac{t_0}{2} \sigma^z (b_0 + b_0^\dagger) +
	\sum_{n=0}^\infty \hbar \omega_n b_n^\dagger b_n  \nonumber \\
	& \hspace*{1.5cm} +\sum_{n=0}^\infty \hbar t_n
	(b_n^\dagger b_{n+1} + b_n b_{n+1}^\dagger)
	\label{eq:H_1D}.
\end{align}
The nearest-neighbor geometry as well as coefficients $\omega_n$ and
$t_n$ are directly related to the recurrence coefficients of the
three-term recurrence relation defining the orthogonal polynomials
$p_n(\omega)$~\cite{chain_mapping_sm}.
\begin{figure}[hbt]
	\begin{center}
	\includegraphics[width=0.9\columnwidth]{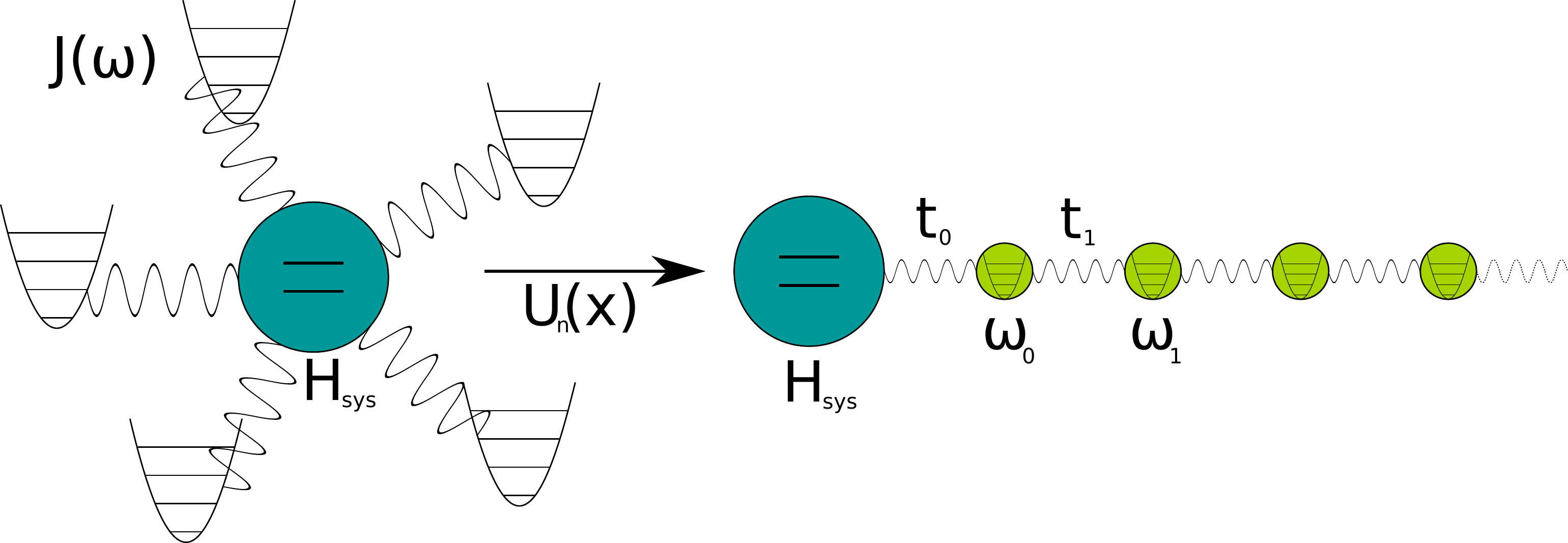}
	\caption{Illustration of the spin-boson model's transformation
	into a one-dimensional configuration where the system is only
	coupled to the environment's first site.}
	\label{fig:tedopa}
	\end{center}
\end{figure}
This transformation from the spin-boson model to a one-dimensional geometry is depicted in Fig.~\ref{fig:tedopa}.

In the second step this emerging configuration is treated by the Time Evolving Block
Decimation (TEBD) method. TEBD  generates a high fidelity approximation of the time evolution of a one-dimensional
system subject to a nearest-neighbor Hamiltonian with polynomially scaling computational resources. TEBD
does so by dynamically restricting the exponentially large Hilbert space
to its most relevant subspace thus rendering the computation feasible~\cite{ref:schollwoeck2011_sm, ref:vidal2004_sm}.
TEBD is essentially a combination of an MPS description for a one-dimensional
quantum system and an algorithm that applies two-site gates that are necessary
to implement a Suzuki-Trotter time evolution. Together with MPS operations such
as the application of measurements this yields a powerful simulation framework.
An extension to mixed states is
possible by introducing a matrix product operator (MPO) to describe the density
matrix, in complete analogy to an MPS describing a state
\cite{ref:schollwoeck2011_sm}. Such an extension is needed in our
simulations in order to build the thermal state of the oscillator chain.

A last step  is necessary to adjust this configuration further to suit
numerical needs. The number of levels for the environment oscillators is restricted to a
value $d_{\text{max}}$ to reduce the required computational
resources. A suitable value for $d_{\text{max}}$ is related to the
sites average occupation which, in turn, depends on the environment structure and
temperature. In our simulations we set $d_{\text{max}}=5$: this value provides converged results
for all examples provided. The Hilbert space dynamical reduction performed by TEBD is
determined to the \emph{bond dimension}. The optimal choice of this parameter depends on the amount of
long range correlations in the system. For all the simulations used in this work, a bond
dimension $\chi=200$ provided converged results.  At last, we observe that the mapping described
above produces a semi-infinite chain that must be truncated in order to enable simulations.  In
order to avoid unphysical back-action on the system
due to finite-size effects, i.e. reflections from the end of the chain, the chain has to be
sufficiently long to completely give the appearance of a ``large'' reservoir. These truncations can
be  rigorously certified by analytical bounds \cite{ref:woods2015_sm}. For the examples provided in the
paper, chains of $n=15$ sites are enough to avoid boundary effects.
In order to further optimize our simulations, we augmented our TEDOPA code with a  Reduced-Rank Randomized Singular Value  Decomposition (RRSVD) routine~\cite{ref:tama2015_sm}.  Singular value decomposition (SVD)
is at the heart of the dimensionality reduction TEBD relies on.  RRSVD is a randomized version of the SVD that provides an improved-scaling SVD, with the same accuracy as the standard state-of-the-art deterministic
SVD routines.

In order to benchmark the quality of the effective model presented in the main text we compared the dynamics of the full spin-boson model in~\eq{eq:H_sb1} with spectral density as in~\eq{jeff_main} of the main text 
with those of a spin coupled to a damped harmonic oscillator in Lindblad description. In the latter case the system evolves according to~\eq{lindblad_main} of the main text with $H=H_{\rm sb1}$ from~\eq{sb_ham_ions_main}
of the main text. As in the main text we chose the parameters $\epsilon = 0$, $\kappa/2\pi=1.25\,$kHz and $\omega_{\rm m}/2\pi=100\,$kHz while we considered a spin-mode
coupling strength $\lambda/2\pi = 100\,$kHz and set the hard cutoff in~\eq{eq:H_sb2} to $\omega_{\rm max}/2\pi = 200\,$kHz. We simulated the dynamics of $\langle \sigma^z (t) \rangle$ for initial product states 
$\ketbra{\uparrow} \otimes \rho_{\beta}$ where $\rho_{\beta}$ is a thermal state at inverse temperature $\hbar \beta = 5.91\cdot 10^{-6}\,$s for the macroscopic environment and a thermal state of a single mode of frequency $\omega_{\rm m}$
with mean occupation $\bar{n} (\omega_{\rm m}) = 0.025$ for the Lindblad case.

\begin{figure}[htb]
\includegraphics[width=\columnwidth]{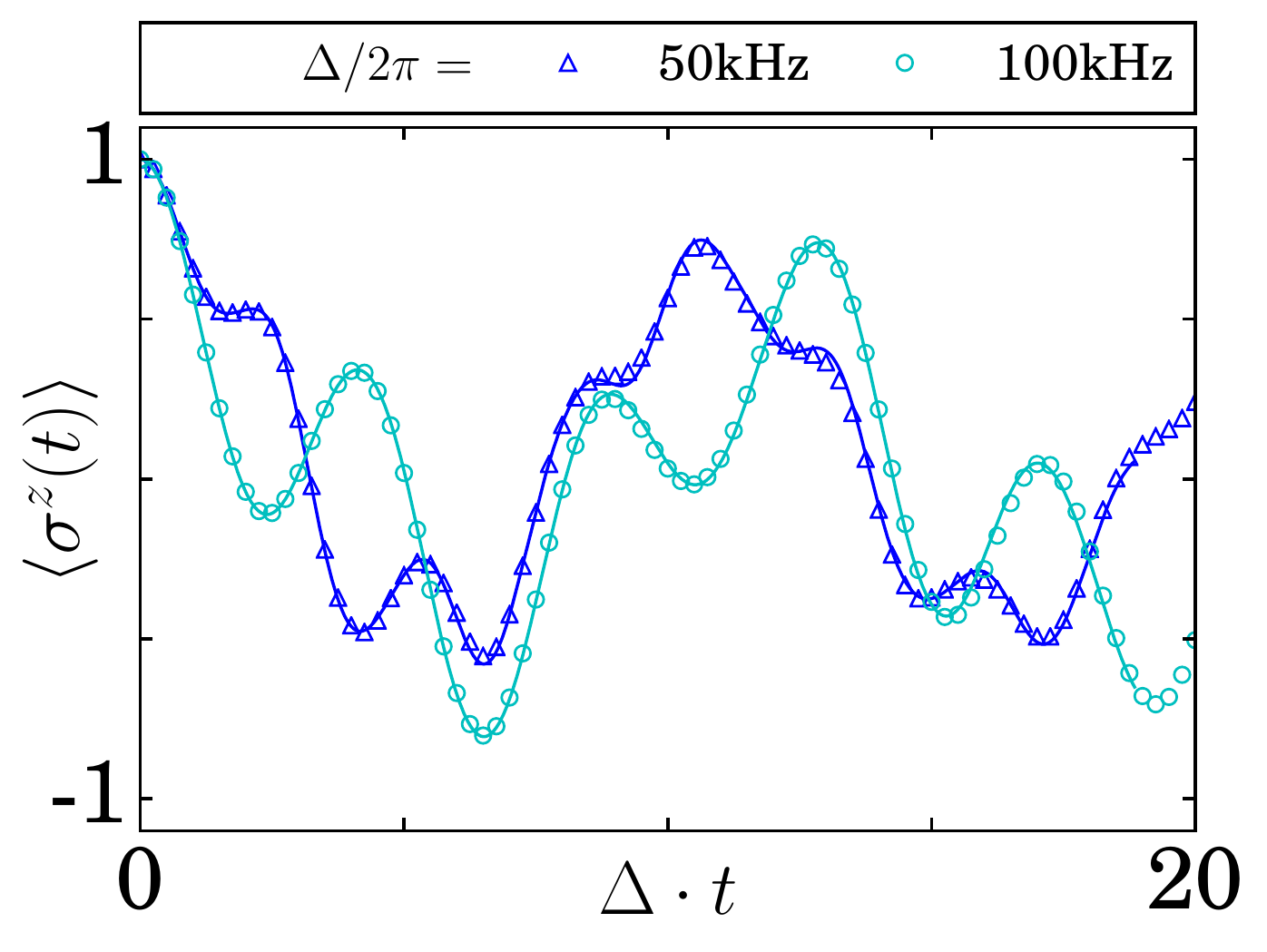}
\caption{The figure shows the dynamics of $\langle \sigma^z (t)\rangle $ for the spin-boson Hamiltonian in~\eq{eq:H_sb1} with spectral density $J_{\rm eff} (\omega)$ as in~\eq{jeff_main} of the main text (solid lines)
and for a spin coupled to a damped mode described by the Lindblad equation~\eqref{lindblad_main} of the main text with the Hamiltonian from~\eq{sb_ham_ions_main} of the main text (triangles and circles). 
The spin energies are $\Delta/2\pi = 100\,$kHz and $\Delta/2\pi = 50\,$kHz. The remaining parameters are given in the text.}
\label{fig_tedopa_comparison}
\end{figure}

The results for spin energies of $\Delta/2\pi = 50,100\,$kHz are shown in Fig.~\ref{fig_tedopa_comparison}. For both cases one can appreciate very good agreement between the two dynamics.
This also shows that the correspondence to the macroscopic environment holds away from the environmental resonance $\omega_{\rm m}/2\pi = 100\,$kHz.

Note that the simulation of one curve for the case $\Delta/2\pi = 50\,$kHz takes 15 days with 16 cores on the bwForCluster JUSTUS such that simulations for the case $\Delta/2\pi = 3\,$kHz presented in the main text are
out of reach.

\section{Spin-dependent optical dipole forces} \label{sec:dipole_forces}

In order to implement the spin-boson Hamiltonian in~\eq{sb_ham_ions_main} of the main text with trapped ions we make use of the so-called spin-dependent optical dipole forces. In this section we derive the Hamiltonian
for the optical dipole forces. For clarity, we consider a somewhat simplified level structure. We employ the formalism of Ref.~\cite{soerensen_reiter_sm} to obtain expressions for the
effective operators of a ground-state manifold weakly coupled to a decaying excited state manifold. 

\begin{figure}[hbt]
\includegraphics[width=.8\columnwidth]{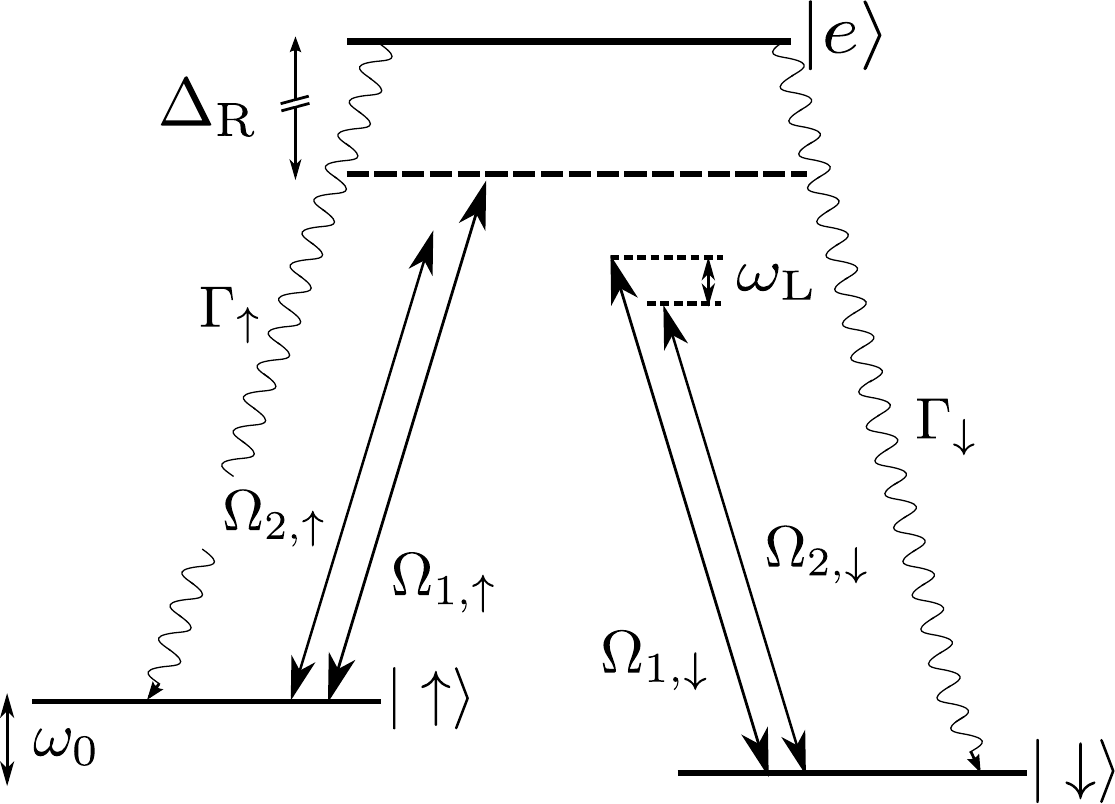}
\caption{The figure shows a three level $\Lambda$-system consisting of the ground states $\ket{\downarrow}$ and $\ket{\uparrow}$ which are separated in frequency by $\omega_0$ and both feature 
a dipole-allowed transition to the decaying excited state $\ket{e}$. The transitions are driven by two lasers and $\Omega_{l,s}$ denotes the Rabi frequency of laser $l$ on transition $\ket{s} \to \ket{e}$.
$\Delta_{\rm R}$ is roughly the detuning of the lasers from the excited state. Depending on the effective laser frequency $\omega_{\rm L} = \omega_1 -\omega_2$ different operations on the ground states 
can be implemented (see text). Spontaneous emission from the excited to the ground states happens at rates $\Gamma_s$ and is indicated by the curly lines.}
\label{lambda_scheme}
\end{figure}

We consider an ion where the internal levels form a $\Lambda$-type three-level system consisting of the ground states $\ket{\uparrow}$ and $\ket{\downarrow}$ which are separated in energy by $\hbar \omega_0$ and have an electric
dipole transition to a decaying excited state $\ket{e}$ (see Fig.~\ref{lambda_scheme}). The free Hamiltonian of the system reads
\beq
H_{\rm at} = \sum_{i=\downarrow, \uparrow, e} \epsilon_i \ketbra{i}
\eeq
with $\epsilon_i$ the energy of the corresponding state. We assume that the dipole transitions are driven by two laser fields with frequencies $\omega_{1/2}$ which couple to both transitions.
In a rotating wave approximation using $|\Omega_{l,s}| \ll \omega_l$ we obtain the interaction Hamiltonian
\beq
H_{\rm L} (t) = \hbar \sum_{l=1,2} \sum_{s=\downarrow, \uparrow} \frac{\Omega_{l,s}}{2} \ee^{ -\ii \omega_{l} t}  \ketbradif{e}{s} + {\rm H.c.}
\eeq
where $\Omega_{l,s}$ is the Rabi frequency of laser $l$ on transition $\ket{s} \to \ket{e}$.
Note that we have included the phase factors $\ee^{\ii({\bf k}_l {\bf r} + \phi_l)}$ where ${\bf r}$ denotes the ion's position and ${\bf k}_l$ ($\phi_l$) the laser wave vector (phase) into the Rabi frequencies.
Finally, we assume that spontaneous emission from the excited level to the ground states is properly described by a dissipator in Lindblad form
\beq
\mathcal{D} \rho= \sum_{s=\downarrow, \uparrow} \left(L_s \rho L_s^{\dagger} - \frac{1}{2}\{L_s^{\dagger} L_s, \rho \} \right)
\eeq
where $L_s = \sqrt{\Gamma_s} \ketbradif{s}{e}$ and $\Gamma =\Gamma_{\uparrow} + \Gamma_{\downarrow} $ is the overall decay rate of the excited state. Putting the pieces together the system evolves according to
\beq
\dot{\rho} = -\frac{\ii}{\hbar}[H_{\rm at} + H_{\rm L}(t), \rho ] + \mathcal{D} \rho.
\eeq

Let us now introduce the detuning
\beq
\delta_{l,s} = (\epsilon_e - \epsilon_s)/\hbar - \omega_l
\eeq
of laser $l$ for transition $\ket{s} \to \ket{e}$. Here, we assume $\delta_{l,s} \simeq \Delta_{\rm R} \gg \omega_0, \Omega_{l,s}, \Gamma$. 
In this case the lasers are far off resonant for all transitions such that the ground states are only weakly coupled to the excited state.
We can then adiabatically eliminate the excited state from the dynamics and obtain an effective dynamics in the ground state manifold. Applying the formalism of~\cite{soerensen_reiter_sm} 
to our system we obtain the effective Lindblad equation
\beq
\dot{\rho} = -\frac{\ii}{\hbar}[H_{\rm eff}, \rho ] + \sum_{k} \left(L^{\rm eff}_k \rho (L^{\rm eff}_k)^{\dagger} 
- \frac{1}{2}\{(L^{\rm eff}_k)^{\dagger} L^{\rm eff}_k, \rho \} \right).
\eeq
The effective Hamiltonian $H_{\rm eff}$ has three contributions $H_{\rm eff} = H_{\rm g} + H_{\rm sr} + H_{\rm odf}$. 
The first part contains the shifted ground state levels
\beq
H_{\rm g} = \sum_{s} (\epsilon_s + \Delta \epsilon_s) \ketbra{s}
\eeq
where the $\Delta \epsilon_s$ are the ac-Stark shifts of the spin-levels due to the applied laser beams
\beq
\Delta \epsilon_s = - \sum_{l,s}\hbar \frac{|\Omega_{l,s}|^2 \delta_{l,s}}{4\delta_{l,s}^2 + \Gamma^2}.
\eeq
The second part, $H_{\rm sr}$, describes two-photon stimulated Raman transitions between the spin states
where a photon is absorbed from one laser beam followed by stimulated emission into the other beam
\beq
H_{\rm sr} = \hbar \sum_{l',l}\frac{\Omega^{\rm sr}_{l',l}}{2} \sigma^+ \ee^{-\ii(\omega_{l'} -\omega_l)t} +{\rm H.c.}
\eeq
Here, we have introduced $\sigma^+ = \ketbradif{\uparrow}{\downarrow} = (\sigma^-)^{\dagger}$ and 
\beq
\Omega^{\rm sr}_{l',l} = -\frac{\Omega_{l,\uparrow}^* \Omega_{l',\downarrow} (\delta_{l',\downarrow} + \delta_{l,\uparrow})}{(2\delta_{l',\downarrow} -\ii \Gamma)(2\delta_{l,\uparrow} +\ii \Gamma)}.
\eeq
The third part of the effective Hamiltonian is a time-dependent ac-Stark shift that can be used to create the optical dipole force
\beq
H_{\rm odf} = \hbar \sum_{s} \frac{\Omega_{s}}{2} \ee^{\ii(\omega_1 -\omega_2)t}\ketbra{s} + {\rm H.c.}
\eeq
where 
\beq
\Omega_{s} = -\frac{\Omega_{1,s}^* \Omega_{2,s} (\delta_{2,s} + \delta_{1,s}) }{(2\delta_{2,s} -\ii \Gamma)(2\delta_{1,s} +\ii \Gamma)}.
\eeq

The Hamiltonian $H_{\rm odf}$ can be written in terms of $\sigma^z = \ketbra{\uparrow}- \ketbra{\downarrow}$ such that we obtain
\beq
H_{\rm odf} = \hbar \frac{\Omega_{\rm rw}}{2} \ee^{-\ii(\omega_1 -\omega_2)t} \mathds{1} + \hbar\frac{\Omega_{\rm odf}}{2} \ee^{-\ii(\omega_1 -\omega_2)t} \sigma^z + {\rm H.c.}
\label{complete_ham_odf}
\eeq
where have introduced the Rabi frequencies
\beq
\Omega_{\rm odf}  = \frac{1}{2}(\Omega_{\uparrow}^* - \Omega_{\downarrow}^*), \hspace{2ex} \Omega_{\rm rw}  = \frac{1}{2}(\Omega_{\uparrow}^* + \Omega_{\downarrow}^*).
\eeq
Thus, we obtain three effects on the spin states. The first is an ac-Stark shift of the spin levels due to the laser fields. The differential 
ac-Stark shift between spin levels can usually be canceled in experiments by adjusting polarization and intensity of the lasers~\cite{pol_and_int_wineland_sm}. Hence, we ignore this contribution. Alternatively,
it could be absorbed into $\omega_0$.

If one chooses the frequency difference between lasers close to the transition frequency between the spin states $\omega_1 -\omega_2 \approx \omega_0$ the second part of the Hamiltonian is resonant and
one can drive coherent two-photon stimulated Raman transitions between the spin states. In this case, we usually have $\Omega_{\rm odf}, \Omega_{\rm rw} \ll \omega_0$, the third contribution
$H_{\rm odf}$ is highly off-resonant and can be neglected in a rotating wave approximation. 

Finally, there is the regime of the spin-dependent optical dipole forces where the beatnote between the two lasers matches one of the motional frequencies $\omega_1 -\omega_2 \approx \omega_k$. 
Usually $\omega_k \ll \omega_0$ such that now the stimulated Raman processes in $H_{\rm sr}$ are highly off-resonant and can be neglected in a rotating wave approximation. Hence, in this regime
we arrive at the effective Hamiltonian
\beq
H_{\rm eff} = \hbar \frac{\omega_0}{2} \sigma^z + \left( \hbar \frac{\Omega_{\rm odf}}{2} \ee^{\ii({\bf k}_{\rm L} {\bf r} + \phi_{\rm L})} \ee^{-\ii \omega_{\rm L} t} \sigma^z + \rm H.c. \right)
\label{eff_ham_odf}
\eeq
with the effective laser frequency $\omega_{\rm L} = \omega_1-\omega_2$ and phase $\phi_{\rm L} = \phi_1 - \phi_2$. Furthermore, we have written the phases $\ee^{\ii {\bf k}_l {\bf r}}$ explicitly again
and introduced the effective laser wave vector ${\bf k}_{\rm L} = {\bf k}_1 - {\bf k}_2$. Note that we have omitted the first part of $H_{\rm odf}$ in~\eq{complete_ham_odf}. For our choice of laser frequency this term would couple
to the motion but it can be canceled choosing the appropriate laser intensities, polarizations and detunings~\cite{pol_and_int_wineland_sm}.
 
Let us turn to the dissipative part. The effective Lindblad operators are found to read:
\begin{eqnarray}
L_{\downarrow}^{\rm eff} = & \sqrt{\Gamma_{\downarrow}} \left(\frac{\Omega_{1,\downarrow} \ee^{-\ii \omega_1 t}}{2\delta_{1,\downarrow} -\ii \Gamma} 
 +\frac{\Omega_{2,\downarrow} \ee^{-\ii \omega_2 t}}{2\delta_{2,\downarrow} -\ii \Gamma} \right) \ketbra{\downarrow} \nonumber \\
& +\sqrt{\Gamma_{\downarrow}} \left(\frac{\Omega_{1,\uparrow} \ee^{-\ii \omega_1 t}}{2\delta_{1,\uparrow} -\ii \Gamma}
+\frac{\Omega_{2,\uparrow} \ee^{-\ii \omega_2 t}}{2\delta_{2,\uparrow} -\ii \Gamma} \right) \ketbradif{\downarrow}{\uparrow}, \\
L_{\uparrow}^{\rm eff} = & \sqrt{\Gamma_{\uparrow}} \left(\frac{\Omega_{1,\uparrow} \ee^{-\ii \omega_1 t}}{2\delta_{1,\uparrow} -\ii \Gamma} 
 +\frac{\Omega_{2,\uparrow} \ee^{-\ii \omega_2 t}}{2\delta_{2,\uparrow} -\ii \Gamma} \right) \ketbra{\uparrow} \nonumber \\
& +\sqrt{\Gamma_{\uparrow}} \left(\frac{\Omega_{1,\downarrow} \ee^{-\ii \omega_1 t}}{2\delta_{1,\downarrow} -\ii \Gamma}
+\frac{\Omega_{2,\downarrow} \ee^{-\ii \omega_2 t}}{2\delta_{2,\downarrow} -\ii \Gamma} \right) \ketbradif{\uparrow}{\downarrow}.
\end{eqnarray}
By keeping only the dominant contributions, i.e. those parts of the action of the Lindblad operators that are time-independent, and using $\delta_{l,s}\simeq \Delta_{\rm R}$ we obtain effective 
operators
\beq
L_{\uparrow\uparrow}= \frac{1}{2}\sqrt{\Gamma_{\uparrow} \sum_l \frac{|\Omega_{l,\uparrow}|^2}{4 \Delta_{\rm R}^2} } \sigma^z, \hspace{2ex}
L_{\downarrow\downarrow}= \frac{1}{2}\sqrt{\Gamma_{\downarrow} \sum_l \frac{|\Omega_{l,\downarrow}|^2}{4 \Delta_{\rm R}^2} } \sigma^z
\label{rayleigh_scattering}
\eeq
and 
\beq
L_{\uparrow\downarrow}= \sqrt{\Gamma_{\uparrow} \sum_l \frac{|\Omega_{l,\downarrow}|^2}{4 \Delta_{\rm R}^2} } \sigma^+, \hspace{2ex}
L_{\downarrow\uparrow}= \sqrt{\Gamma_{\downarrow} \sum_l \frac{|\Omega_{l,\uparrow}|^2}{4 \Delta_{\rm R}^2} } \sigma^-.
\label{raman_scattering}
\eeq
The first two terms describe Rayleigh scattering where the spin state is not altered upon a scattering event but can introduce dephasing. The other operators describe Raman scattering
where the spin state is changed upon a scattering event. If we assume the modulus of the Rabi frequencies is approximately equal 
$| \Omega_{l,s} | \approx \Omega_0$, we can estimate the effective scattering rate $\Gamma_{\rm eff} \approx \Gamma \Omega_{\rm L}/\Delta_{\rm R}$ where $\Omega_{\rm L} = \Omega_0^2/(2\Delta_R)$
is the approximate effective laser Rabi frequency. Hence, decoherence can be largely suppressed if we choose $\Delta_{\rm R}$ large enough.

\section{Spin-boson Hamiltonian with trapped ions} \label{sec: spin-boson with ions}

In this section we want to show how to obtain the spin-boson Hamiltonian in~\eq{sb_ham_ions_main} of the main text in an ion trap experiment. For definiteness we chose to consider a 
$^{24}{\rm Mg}^+ - ^{25}{\rm Mg}^+$ crystal. $^{25} {\rm Mg}^+$ has electronic hyperfine ground states with total angular momentum $F=2,3$ for the valence electron in the $^2 S_{1/2}$ state
whose degeneracy can be lifted by a magnetic field. A possible choice for a qubit are the states $\ket{F= 3, m_F = 3} \equiv \ket{\downarrow} $ and $\ket{F= 2, m_F = 2} \equiv \ket{\uparrow} $.
The hyperfine splitting between the $F=2$ and $F=3$ states is about $\omega_0/2\pi \simeq 1.8\,$GHz. At a magnetic field $B= 4\,$G the other hyperfine states are well-separated from the qubit states
due to the Zeeman interaction and we can assume the Hamiltonian
\beq
H_{\rm s} = \hbar\frac{\omega_0}{2} \sigma^z
\eeq
for the internal levels of $^{25} {\rm Mg}^+$ where $\sigma^z = \ketbra{\uparrow} - \ketbra{\downarrow}$.

The two ions interact through their Coulomb interaction and their motion is coupled. If the ions are sufficiently cold they form a so-called Coulomb crystal and
perform only small oscillations about equilibrium. We assume trapping conditions such that the ions form a string along $z$ and their equilibrium positions read
${\bf r}_j^0 = (0,0,z_j^0)^T$. Their motion is then conveniently 
described in terms of normal modes~\cite{james_hom_crystals_sm, morigi_inhom_crystals_sm}. 
For a crystal of $N$ ions we obtain $N$ modes in each direction such that, taking into account the coupled harmonic motion, the system's Hamiltonian becomes
\beq
H_0 = \hbar\frac{\omega_0}{2} \sigma^z + \sum_{\alpha,n} \hbar\omega_{\alpha,n} a_{\alpha,n}^{\dagger} a_{\alpha,n}.
\eeq
Here, $\omega_{\alpha,n}$ is the frequency of mode $n$ in direction $\alpha$ and $a_{\alpha,n}^{\dagger}\:(a_{\alpha,n})$ creates (annihilates) an excitation in the corresponding mode.
$^{24} {\rm Mg}^+$ is used to sympathetically cool the ions' coupled motion. Since the internal levels are adiabatically eliminated in the description of laser cooling~\cite{cirac_laser_cooling_sm, morigi_eit_cooling_sm}
we have omitted them here. The spin transition can be driven either directly by a microwave or in a two-photon stimulated-Raman configuration (see previous section). We adopt the convention that we will call the
field driving the spin transition the ``microwave'' independent of the physical realization.

Let us now assume the spin is driven by a microwave with frequency $\omega_{\rm d}$ and Rabi frequency $\Omega_{\rm d}$ and we apply a spin-dependent force as in~\eq{eff_ham_odf}. 
The interaction Hamiltonian then reads
\beq
H_{\rm int} = \hbar\frac{\Omega_{\rm d}}{2} \sigma^+ \ee^{-\ii \omega_{\rm d} t} + \hbar\frac{\Omega_{\rm odf}}{2} \ee^{ \ii ({\bf k}_{\rm L} {\bf r} + \phi_{\rm L})} \ee^{- \ii \omega_{\rm L} t} \sigma^z + \rm H.c. 
\eeq
where we have set the microwave phase to zero and performed a rotating wave approximation. $\Omega_{\rm odf}$ denotes the effective laser Rabi frequency, %given in~\eq{odf_rabi_freq} and 
$\omega_{\rm L}$, ${\bf k}_{\rm L}$ and $\phi_{\rm L}$ the effective laser frequency, wave vector and phase. We assume ${\bf k}_{\rm L} =k {\bf e}_z $ such that the laser only couples to the motion along $z$. 
We have ${\bf r}_{jz}= z_j^0 + z_j$ where the $z_j$ can be written in terms of the quantized normal modes~\cite{morigi_inhom_crystals_sm}:
\beq
z_j = \sum_n \tilde{M}_{jn}\sqrt{\frac{\hbar}{2 m_j \omega_n}} (a_n + a_n^{\dagger})
\eeq
where $m_j$ is the mass of ion $j$, $\tilde{M}_{jn}$ the amplitude of motional mode $n$ at ion $j$ in mass-weighted coordinates and $\omega_n = \omega_{z,n}$ (for the operators accordingly).
The full Hamiltonian of the system then reads
\beq
H= H_0 + H_{\rm int}.
\eeq
Moving to an interaction picture with respect to $\tilde{H}_0 = \hbar(\omega_{\rm d}/2) \sigma^z + \hbar\sum_{\alpha,n} \omega_{\alpha,n} a_{\alpha,n}^{\dagger} a_{\alpha,n}$ we obtain
the transformed interaction Hamiltonian
\beq
\begin{split}
&\tilde{H}_{\rm int} = \hbar\frac{\delta}{2} \sigma^z + \hbar\frac{\Omega_{\rm d}}{2} \sigma^x \\
&+ \left(\hbar \frac{\tilde{\Omega}_{\rm odf}}{2} \ee^{ \ii \sum_n \eta_n (a_n \ee^{- \ii \omega_{n} t} + a_n^{\dagger} \ee^{ \ii \omega_{n} t})} \ee^{- \ii \omega_{\rm L} t} \sigma^z + \rm H.c. \right)
\end{split}
\eeq
where $\delta = \omega_0 - \omega_{\rm d}$, $\tilde{\Omega}_{\rm odf} =\Omega_{\rm odf} \ee^{\ii(k z_2^0 + \phi_{\rm L})}$ and we have introduced the Lamb-Dicke factors 
$\eta_n = \tilde{M}_{2n}k\sqrt{\hbar/(2 m_2 \omega_n)}$. Note that we have assumed that the $^{25}{\rm Mg}^+$ ion is located at site~2.

Usually for an optical wave vector $\eta_n \ll 1$ such that we can expand the exponential to first order in the $\eta_n$. In the axial direction the two-ion crystal features an in- and out-of-phase
mode of motion that are well separated in frequency. More precisely, we consider a trapping potential such that a single $^{24}{\rm Mg}^+$ has a center-of-mass frequency 
$\omega_{\rm m}/2\pi = 2.54\,$MHz. The in- and out-of-phase mode frequencies of the $^{24} {\rm Mg}^+ - ^{25} {\rm Mg}^+$ crystal are then given by $\omega_{1}/2\pi=2.51\,$MHz 
and $\omega_{2}/2\pi=4.36\,$MHz , respectively. 
If we choose the laser frequency close to the out-of-phase mode frequency $\omega_{\rm L} \approx \omega_2$ and $\Omega_{\rm odf} \ll 2 \omega_{\rm L},\eta_1 \Omega_{\rm odf} \ll |\omega_1 -\omega_{\rm L}|$
we can neglect all terms except the coupling to the out-of-phase mode in a rotating wave approximation and arrive at the final Hamiltonian
\beq
\tilde{H}_{\rm int} = \hbar\frac{\delta}{2} \sigma^z + \hbar\frac{\Omega_{\rm d}}{2} \sigma^x + 
\left(\hbar \ii \eta_2 \frac{\tilde{\Omega}_{\rm odf}}{2} a_2^{\dagger} \sigma^z \ee^{\ii \delta_{\rm m} t} + {\rm H.c.}\right)
\eeq
where $\delta_{\rm m} = \omega_2 - \omega_{\rm L} \ll \omega_2$ is the detuning of the laser from the out-of-phase mode and we choose $\omega_{\rm L}$ such that $\delta_{\rm m} > 0$. Finally,
we can cast the above Hamiltonian in a time-independent form and we recover~\eq{sb_ham_ions_main} of the main text
\beq
\tilde{H}_{\rm int}/\hbar = \frac{\delta}{2} \sigma^z + \frac{\Omega{\rm d}}{2} \sigma^x - 
\frac{\lambda}{2} (a_2 + a_2^{\dagger})\sigma^z + \omega_{\rm m} a_2^{\dagger} a_2
\eeq
where $\lambda = -\ii\eta_2 \tilde{\Omega}_{\rm odf}$ can always be taken to be real and $\delta_{\rm m} =\omega_{\rm m}$. Thus, the mode frequency in our simulation is given by the detuning of the 
spin dependent force. Making the substitutions $\hbar \delta = \epsilon$ and $\Omega_{\rm d} = -\Delta$  we obtain the spin-boson Hamiltonian for a single mode. 

Note that experimentally a finite bias $\epsilon$ can easily be included by introducing a detuning to the field driving the spin transition. For the spin-motion coupling we consider one has to take care
that the laser beams providing the spin-motion coupling are sufficiently detuned such that the simulation is not compromised by errors due to photon scattering (see previous section). In order to avoid this source of error
one could also rotate the spin basis and provide spin-motion coupling e.g. by a M{\o}lmer-S{\o}rensen interaction~\cite{ms_gates_sm}.

\section{Computation of non-Markovianity measures}

There are several different ways to define non-Markovian dynamics. Here, we start by reviewing the definition presented in~\cite{rhp_review_sm}. Let us consider a quantum system
whose time evolution is described by a completely positive and trace preserving dynamical map $\mathcal{E}_{t,t_0}$. Then for an initial state $\rho (t_0)$ the system's state at
a later time $t \geq t_0$ is given by
\beq
\rho(t) = \mathcal{E}_{t,t_0} \rho(t_0).
\eeq 
According to~\cite{rhp_review_sm} the dynamical map describes a Markovian evolution if and only if the map $\mathcal{E}_{t_2,t_1}$ exists and is completely positive for all 
$t_2 \geq t_1 \geq t_0$ .The degree of non-Markovianity of a dynamics over an interval $I$, $\mathcal{N}_{\rm RHP}$, is then obtained by quantifying the departure of 
$\mathcal{E}_{t_2,t_1}$ from complete positivity over that interval. In particular, we have
\beq
\mathcal{N}_{\rm RHP} = \frac{\int_{I,\bar{g}>0} \bar{g}(t) {\rm d} t }{ \int_{I,\bar{g}>0} \chi [\bar{g}(t)] {\rm d} t}
\label{nrhp_sm}
\eeq
where the integral extends over those subintervals of $I$ where $\bar{g}(t) >0$. The function $\chi[x] = 1$ for $x>0$ and $\chi[x] = 0$ else and
by definition ``$0/0$''$ = 0$ . The function $\bar{g}(t)$ is given by $\bar{g}(t)=\tanh[g(t)]$ where
\beq
g(t) = \lim_{\epsilon \to 0^+} \frac{ \| [\mathcal{E}_{t+\epsilon,t}\otimes \mathds{1}] \ketbra{\psi} \|_1-1}{\epsilon}
\eeq
where $\| \dots \|_1$ denotes the trace norm and $\ket{\psi} = \frac{1}{\sqrt{d}}\sum_{n=1}^{d} \ket{n,n}$ is a maximally entangled state of the open system of finite dimension $d$
with an ancillary system of the same size. Note that we restrict our considerations to finite dimensional open systems. $[\mathcal{E}_{t+\epsilon,t}\otimes \mathds{1}] \ketbra{\psi}$ 
is the so-called Choi matrix and is positive if and only if $\mathcal{E}_{t+\epsilon,t}$ is completely positive. Note that $g(t)$ vanishes if $\mathcal{E}_{t+\epsilon,t}$
is completely positive. Thus, for a Markovian dynamics $g(t)=0$ for all times and $\mathcal{N}_{\rm RHP}$ evaluates to zero.

We evaluated $\mathcal{N}_{\rm RHP}$ numerically for the spin-boson system consisting of a spin coupled to a damped mode described by~\eq{lindblad_main} of the main text with the Hamiltonian
in~\eq{sb_ham_ions_main}.
To this end we divide the time interval $I=[0,T]$ that we want to inspect for non-Markovian dynamics in $N$ equally spaced discrete times $t_i$ ($t_0=0,t_N=T$) and compute the
time evolution of the basis states $\ketbradif{k}{j},\:k,j=\uparrow,\downarrow$ for all $t_i$. By writing the time-evolved states $\ketbradif{k}{j}(t_i) = \rho_{kj} (t_i)$ as
a vector $v_{kj}(t_i) = [\rho_{kj,\uparrow \uparrow} (t_i),\rho_{kj,\uparrow \downarrow} (t_i),\rho_{kj,\downarrow\uparrow}(t_i),\rho_{kj,\downarrow\downarrow}(t_i)]^T$ 
we can write the dynamical map $\mathcal{E}(t,t_0)$ in matrix representation
\beq
E(t,t_0) = [v_{\uparrow\uparrow}(t),v_{\uparrow\downarrow}(t),v_{\downarrow\uparrow}(t),v_{\downarrow\downarrow}(t)].
\eeq 
The matrix for the time evolution from $t_1$ to $t_2$ where $t_2 \geq t_1 \geq t_0$ is then computed by
\beq
E(t_2,t_1) = E(t_2,t_0) E^{-1}(t_1,t_0)
\eeq
where $E^{-1}(t_1,t_0)$ is the normal matrix inverse. The Choi matrix $[\mathcal{E}_{t_2,t_1}\otimes \mathds{1}] \ketbra{\psi}$ is proportional to the reshuffled matrix $E^{\rm R} (t_2,t_1)$
of the matrix $E (t_2,t_1)$~\cite{bengtsson_sm}. In particular, the Choi matrix is given by
\beq
[\mathcal{E}_{t_2,t_1}\otimes \mathds{1}] \ketbra{\psi} = \frac{1}{d} E^{\rm R} (t_2,t_1)
\eeq
where $d$ is the dimension of the finite dimensional open quantum system. For the case of a spin $E^{\rm R} (t_2,t_1)$ reads
\beq
E^{R} (t_2,t_1) = \begin{pmatrix} 
E_{11} & E_{12} & E_{21} & E_{22} \\
E_{13} & E_{14} & E_{23} & E_{24} \\
E_{31} & E_{32} & E_{41} & E_{42} \\
E_{33} & E_{34} & E_{43} & E_{44} 
\end{pmatrix}.
\eeq
where $E_{mn}$ corresponds to entry $m,n$ of the $4 \times 4$ matrix $E (t_2,t_1)$. Now, in order to obtain $\mathcal{N}_{\rm RHP}$ we evaluated a discrete version of $g(t)$ 
according to 
\beq
g(t_i) = \frac{ \| [\mathcal{E}_{t_{i+1},t_i} \otimes \mathds{1}] \ketbra{\psi} \|_1-1}{t_{i+1}-t_i} = \frac{ \| \frac{1}{d} E^{\rm R} (t_{i+1},t_i) \|_1-1}{t_{i+1}-t_i}.
\eeq
The difficulty in evaluating $g(t_i)$ is to decide which values of the numerator count as zero and which are counted as finite. The numerical calculations were performed
using Python's Numpy and Scipy libraries. The oscillator's Hilbert space was truncated at a maximal phonon number $n_{\rm max} = 15$. The states were evolved in time by vectorizing the
Lindblad equation and applying the matrix exponential of the Liouvillian on the vectorized form of the density matrix using the scipy.sparse.linalg.expm\_multiply routine. For a number of parameters 
the resulting density matrices were compared to the density matrices obtained by performing the matrix exponential first with scipy.sparse.linalg.expm and then the matrix vector multiplication.
For all of the spin basis states the resulting matrices typically showed trace distances of a few times $10^{-16}$. Summing the largest errors of all the basis states yielded a few times $10^{-15}$. Taking this value as a rough estimate
of the numerical precision we set $g(t)=0$ if the numerator was smaller than $10^{-14}$. Finally, $\mathcal{N}_{\rm RHP}$ in this numerical approximation is given by 
\beq
\mathcal{N}_{\rm RHP} = \frac{\sum_{i=1,g(t_i)>0}^{N} \tanh[{g}(t_i)]  }{ N_{g(t_i)>0}}
\label{nrhp_sm}
\eeq
where $N_{g(t_i)>0}$ is the number of events where $g(t_i)>0$. For the ``ohmic'' case ($\Delta/2\pi = 3\,$kHz) we chose $T=0.01/\Delta$ and $N=10^4$ and for the resonant case
 ($\Delta/2\pi = 100\,$kHz)  $T=0.1/\Delta$ and $N=10^4$. Note that taking a too small time steps eventually leads to discontinuous behavior in $\mathcal{N}_{\rm RHP}$.
 
The computation of the measure of non-Markovianity $\mathcal{N}_{\rm BLP}$~\cite{blp_review_sm} is somewhat easier.  
$\mathcal{N}_{\rm BLP}$ was originally proposed as a measure of non-Markovianity based on the monotonicity of the trace distance under completely positive and trace preserving
evolutions and is given by
\beq
\mathcal{N}_{\rm BLP} = \max_{\rho_{1/2}} \int_{I,\sigma > 0 } \sigma(t) {\rm d}t
\label{nblp_sm}
\eeq
where $\sigma(t) = \frac{\rm d}{{\rm d}t} D(\mathcal{E}_{t,t_0}\rho_1,\mathcal{E}_{t,t_0}\rho_2)$ and $D(\cdot,\cdot)$ is the trace distance.
The integral extends over those subintervals of $I$ where $\sigma (t) > 0$. Thus, $\mathcal{N}_{\rm BLP}$ detects non-Markovianity of a dynamical map $\mathcal{E}_{t,t_0}$
if the trace distance between two initial states $\rho_1$ and $\rho_2$ increases in the course of the dynamics induced by $\mathcal{E}_{t,t_0}$. 
A nonzero value of $\mathcal{N}_{\rm BLP}$ can be associated with a backflow of information from the environment to the system~\cite{blp_review_sm}.
It is known that optimal state pairs $\rho_1,\:\rho_2$ that saturate the maximum in~\eq{nblp_sm} are orthogonal and lie on the boundary of state space~\cite{optimal_states_nblp_sm}.
However, since we only want to witness non-Markovian dynamics we do not need to perform the maximization in~\eq{nblp_sm}.
Therefore, we can provide a useful lower bound on $\mathcal{N}_{\rm BLP}$ by computing the measure for the eigenstates $\ket{\uparrow/\downarrow}$,
$\ket{\pm}_x$ and $\ket{\pm}_y$ of the Pauli matrices $\sigma^z,\:\sigma^x$ and $\sigma^y$, respectively.

For the numerical computation of $\mathcal{N}_{\rm BLP}$ we considered the whole interval $[0,20/\Delta]$. As in the previous case we considered $N=10^4$ equally spaced points
$t_i$ in the interval and computed the time evolution for the spin starting in each of the eigenstates of the Pauli matrices. We then computed the discrete version of 
$\mathcal{N}_{\rm BLP}$
\beq
\mathcal{N}_{\rm BLP} = \sum_{i,D_{t_{i+1}} - D_{t_{i}} > 0} (D_{t_{i+1}} - D_{t_{i}}) 
\label{nblp_discrete}
\eeq
for each pair of eigenstates. Here the sum runs over those $i$ where the term in brackets is larger than zero and $D_{t_{i}} =D(\mathcal{E}_{t_i,t_0}\rho_1,\mathcal{E}_{t_i,t_0}\rho_2)$.
We note that due to the finite number of ``measurements'' there will be small deviation to the true value of $\mathcal{N}_{\rm RHP}$~\cite{nblp_measurement_sm}.
The values shown in Fig. 2 of the main text are obtained for the initial spin states $\rho_{\rm s}(0) = \ketbra{\pm }_x$ in the Ohmic case and $\rho_{\rm s} (0)= \ketbra{\uparrow},\ketbra{\downarrow}$
in the resonant case.

\end{document}